\documentclass[onecolumn,amsmath,amssymb,pre]{revtex4-2}

\usepackage{graphicx}
\usepackage{color}
\usepackage{dcolumn}
\usepackage{amsmath}
\usepackage{subfigure}

\usepackage{sidecap}
\usepackage{dsfont}

\usepackage{mathrsfs}
\usepackage{amsfonts}
\usepackage{indentfirst}
\usepackage{bm}
\usepackage{tikz}
\usepackage{ORCIDinREVTeX}

\usepackage[colorlinks,citecolor=blue]{hyperref}

\newcommand{\be}{\begin{equation}}
\newcommand{\ee}{\end{equation}}
\newcommand{\bey}{\begin{eqnarray}}
\newcommand{\eey}{\end{eqnarray}}
\newcommand{\bw}{\begin{widetext}}
\newcommand{\ew}{\end{widetext}}

\newcommand{\ra}{\rangle}
\newcommand{\la}{\langle}

\newcommand{\ba}{\begin{array}}
\newcommand{\ea}{\end{array}}
\newcommand{\bi}{\begin{itemize}}
\newcommand{\ei}{\end{itemize}}
\newcommand{\bem}{\begin{enumerate}}
\newcommand{\eem}{\end{enumerate}}

\newcommand{\nint}[1]{\ensuremath\left\lfloor#1\right\rceil}

\begin{document}

\title{Characterizing the mixed eigenstates in kicked top model 
through the out-of-time-order correlator}

\author{Qian Wang}
\orcid{0000-0002-3937-5657}
\affiliation{CAMTP-Center for Applied Mathematics and Theoretical Physics, University of Maribor, 
Mladinska 3, SI-2000 Maribor, Slovenia, and \\
Department of Physics, Zhejiang Normal University, Jinhua 321004, China}
\author{Marko Robnik}
\orcid{0000-0002-1098-1928}
\affiliation{CAMTP-Center for Applied Mathematics and Theoretical Physics, University of Maribor, 
Mladinska 3, SI-2000 Maribor, Slovenia}

\begin{abstract}

Generic systems are associated with a mixed classical phase space.
The question of the properties of the eigenstates for these systems 
remains less known, although it plays a key role for understanding
several important quantum phenomena such as 
thermalization, scarring, tunneling, and (de-)localization.
In this work, by employing the kicked top model, we perform a detailed investigation of the 
dynamical signatures of the mixed eigenstates via the out-of-time-order correlator (OTOC).
We show how the types of the eigenstates get reflected in
the short- and long-time behaviors of the OTOC and conjecture that  
the dynamics of the OTOC can be used as an indicator of the mixed eigenstates. 
Our findings further confirm the usefulness of the OTOC 
for studying quantum complex systems and also 
provide more insights into the characters the mixed eigenstates.

\end{abstract}

\date{\today}

\maketitle

\section{Introduction}

Studying dynamical properties of eigenstates in quantum many-body systems 
has attracted much attention in modern science, due to their 
crucial role for understanding numerous fundamental questions arisen in different research areas,
including quantum chaos \cite{Das2023,ShiZ2023,Luca2016,Herrera2019,Pandey2020}, 
statistical mechanics \cite{Herrera2019,Luca2016}, and condensed matter physics \cite{Nandkishore2015,Abanin2019}. 
Moreover, the endeavor to investigate the dynamical features of eigenstates in quantum systems
is also pivotal for various applications of quantum-based techniques, 
such as quantum simulation \cite{Georgescu2014,Daley2022} 
and metrology \cite{Giovannetti2011,Magdalena2016}.

Numerous works have been devoted to investigating the dynamical 
signatures of eigenstates in the fully chaotic systems, 
see e.~g. Refs.~\cite{Pandey2020,Herrera2019,Luca2016} and references therein. 
However, a generic many-body system is neither regular nor fully chaotic. 
Instead, it behaves as a mixed-type system.
Pretty much different from both regular and strong chaotic cases,
the mixed-type systems are characterized by 
the coexistence of regular islands and chaotic regions 
in phase space of their classical counterparts.
Hence, the quest to classify the eigenstates in mixed-type systems
is important for studying them.    

Percival in his seminal work \cite{Percival1973} proposed to classify the 
spectra of quantum mixed-type systems into
regular and chaotic eigenstates, supported, respectively, 
by invariant tori in regular islands and chaotic seas 
in corresponding classical phase space \cite{Stechel1984}.
This proposal has been further developed by 
Berry an coworkers \cite{Berry1977,Victor1977} 
and finally leads to the so-called principle of uniform 
semiclassical condensation (PUSC) of Wigner (or Husimi) functions \cite{Robnik2000}. 
This is the basis for the Berry-Robnik picture regarding the statistical properties of the energy spectra, 
the level spacings distribution \cite{BerryR1984}.
See Refs.~\cite{Robnik2020,Robnik2024} and references therein for more details about the PUSC.  

Although the binary separation of quantum eigenstates has been verified 
and commonly accepted in the studies of quantum chaos, 
the picture for reality situations is more complicated. 
Actually, the sharp distinction between chaotic and regular 
eigenstates only happens in the ultimate semiclassical limit. 
Moreover, it is known that different phase space structures of mixed-type systems can be connected 
through various tunneling processes \cite{Tomsovic1994,Frischat1998,Lock2010}. 
These facts strongly indicate that the mixed-type systems 
also allow their eigenstates behaving as mixed states.
In contrast to the regular and fully chaotic eigenstates,
the Husimi function of mixed eigenstates is distributed 
in both regular and chaotic regions \cite{Lozej2022,WangR2023,WangR2024}.
This raises a natural and intriguing question: what are the properties of the mixed eigenstates?
Previous works have examined their statistical properties \cite{Varma2024}
and how their relative fraction decreases as the 
semiclassical limit is approached \cite{Lozej2022,WangR2023,WangR2024,YanR2024,YanWR2024} 
in several mixed-type systems.  
However, a detailed understanding of their dynamical features is still lacking in current studies. 

In the present work, we make a step toward addressing this question.
To this end, we carry out a thorough analysis of the dynamical 
signatures of the mixed eigenstates in quantum kicked top model by means of the 
out-of-time-order correlators (OTOCs). 
As a measure of quantum information scrambling \cite{XuSheng2019,Gonzalez2019,YanB2020,XuSheng2024}, 
the concept of OTOC was first introduced for the semiclassical 
study of superconductivity \cite{Larkin1969}
and, recently, it has been commonly used in 
condensed matter \cite{ChenX2017, RFan2017,Dora2017,LinC2018} 
and high-energy physics \cite{Shenker2014,Maldacena2016}.
In particular, they exhibit an initial exponential growth behavior for the 
quantum systems with chaotic classical counterpart \cite{Rozenbaum2017,Garcia2018,Carlos2019,Lewis2019},
leading to the so-called quantum butterfly effect \cite{Aleiner2016,Cotler2018}.  
As a result, the OTOCs have been recognized as the quantum analogoue  
of classical instability with respect to initial condition. 
This triggers a vast amount of studies on the connections between 
OTOCs and quantum chaos 
\cite{Fortes2019,XuT2020,Kirkby2021,Dowling2023,Rozenbaum2019,Rautenberg2020,
Alonso2022,Trunin2023,Novotny2023,Garcia2023,Shukla2024}. 
However, it should be emphasized that such exponential instability occurs only at finite time (i.~e. Ehrenfest time),
unlike the classical case with positive Lyapunov exponents. 
Moreover, the experimental measurement of OTOCs has been accomplished by several platforms \cite{LiJun2017,Garttner2017,Braumuller2022,Green2022,Blocher2022}.

The main interest of this work is to explore how the mixed eigenstates 
get manifested in the evolution of OTOC.
We thus focus on the eigenstate expectation values of OTOC, 
which enables us to analyze the dependence of the behavior of OTOC on the type of eigenstates.
We demonstrate that the mixed feature of the eigenstates results in strong 
impact on the evolution of OTOC and discuss how to reveal the dynamical 
signatures of mixed eigenstates via the properties of OTOC. 
Specifically, we show that the phase space overlap of the mixed eigenstates correlates with 
both short-time growth rate and long-time average of the OTOC.  

The rest of the article is structured as follows. 
In Sec.~\ref{Second}, we introduce some basic concept of the OTOCs; 
we provide a short review of the kicked top model, 
and briefly recall the definition and characterization of the mixed eigenstates. 
Afterwards, we will report and discuss our results in Sec.~\ref{Third}, wherein 
we show how the mixed eigenstates affect the dynamics of the OTOC.  
We finally draw our conclusions in Sec.~\ref{Fourth}.

 \begin{figure}
  \includegraphics[width=\textwidth]{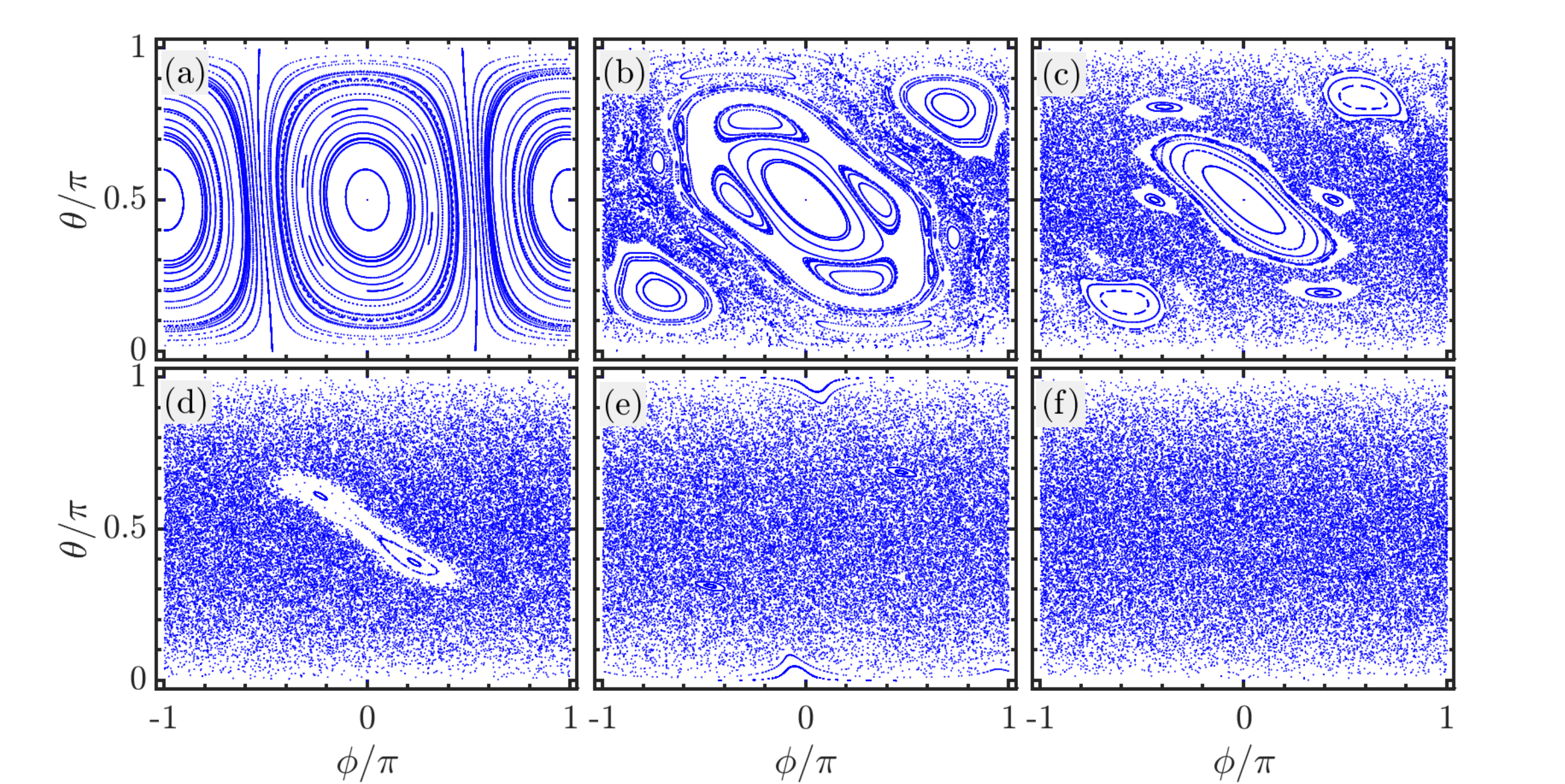}
  \caption{Classical phase space portraits of the kicked top model
  for $121$ random initial conditions with 
  $\kappa=0.2, 2.2, 3, 4.3, 6$, and $\kappa=7$ [from (a) to (f)].
  Each initial condition has been evolved for $300$ kicks. 
  Other parameter: $\alpha=13\pi/19$.}
  \label{PSections}
 \end{figure}

\section{Backgrounds} \label{Second}

In this section, we introduce the OTOCs and the model studied in this work.
We also briefly discuss the definition and characterization of the mixed eigenstates, 
the main topic of the present work.

\subsection{Out-of-time-order correlators}

The OTOCs quantify the information scrambling 
in quantum many-body systems \cite{XuSheng2024}. 
They have attracted a great deal of attention from both theoretical
\cite{Lashkari2013,Kukuljan2017,Schuster2023} 
and experimental \cite{LiJun2017,Garttner2017,Braumuller2022} aspects in recent years.   
For a system evolved according to the Hamiltonian $H$, the OTOC 
of two Hermitian operators $A$ and $B$ is defined as
\be
   C(t)=\la[A(t), B]^\dag[A(t),B]\ra=-\la[A(t),B]^2\ra,
\ee
where $A(t)=U^\dag(t) AU(t)$ with $U(t)=e^{-iHt}$ being the time evolution operator.
Here, we set $\hbar=1$ throughout this work and $\la\cdots\ra$ 
denotes an average for certain quantum state. 
The OTOCs are usually evaluated as thermal average over the 
canonical ensemble with inverse temperature $\beta$
\cite{Maldacena2016,Hashimoto2017}.
However, in order to reveal the dynamical property of a single state, 
we consider $C(t)$ as an expectation value
for a fixed eigenstate of the system, namely the so 
called microcanonical OTOC \cite{Hashimoto2017,Novotny2023,Carlos2019}.

 \begin{figure}
  \includegraphics[width=\textwidth]{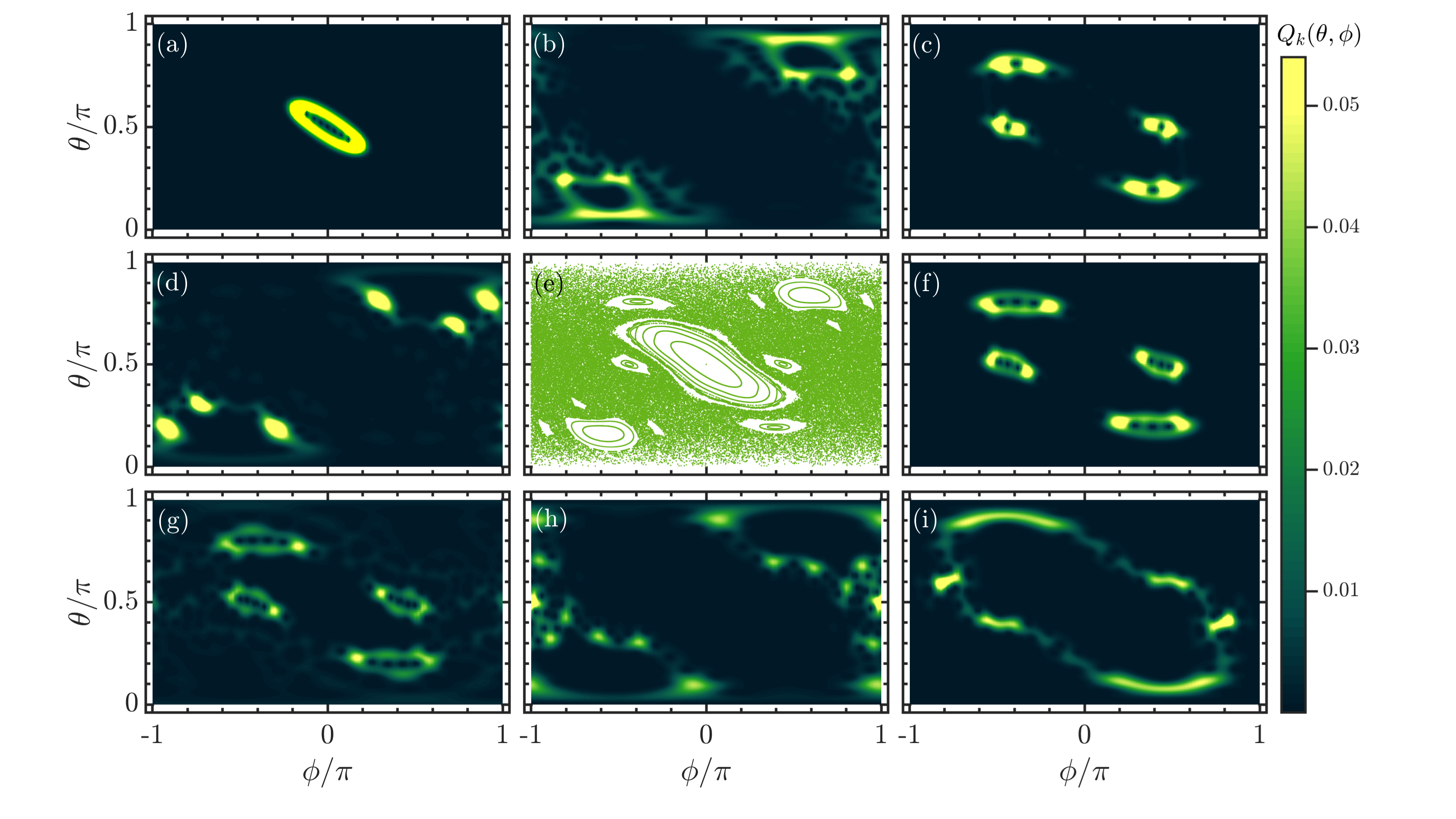}
  \caption{Husimi function $Q_k(\theta,\phi)$ in (\ref{HsFD}) for several 
  eigenstates of the Floquet operator with phase space 
  overlap indices are: (a) $\chi_{6}=-1$, (b) $\chi_{85}=0.6371$, (c) $\chi_{113}=-0.275$,
  (d) $\chi_{147}=0.5072$, (f) $\chi_{179}=0.3858$, (g) $\chi_{208}=0.7262$, (h) $\chi_{230}=0.9780$, 
  and (i) $\chi_{273}=0.9985$. 
  The corresponding classical phase portrait is shown in panel (e). 
  Other parameters: $\alpha=13\pi/19$, $\kappa=3$, and the system size $S=150$.}
  \label{Husimifs}
 \end{figure}

Although the quantum-classical correspondence \cite{Larkin1969,Cotler2018} 
implies that a general OTOC increases exponentially with time until the
well-known Ehrenfest (or scrambling) time for the 
quantum system with classical chaotic counterpart
\cite{Carlos2019,Rozenbaum2017,Garcia2023},
whether the OTOC can be used as a dynamical indicator of 
quantum chaos is still under debate 
\cite{Yan2019,Cameo2020,Rozenbaum2020,XuT2020,Hashimoto2020,WangJ2021,Kirkby2021,Dowling2023}.  
Nonetheless, the OTOC analysis could provide more insights 
into the dynamical features of both isolated  
\cite{Lakshmin2019,Herrera2018, Rammensee2018, Bergamasco2019,Borgonovi2019,
Riddell2023,Balachandran2023,Varikuti2024}
and open \cite{Syzranov2018,Chatterjee2020,ZhaiL2020,ZhaoW2022,Bergamasco2023}
quantum systems.
OTOC is also widely used for studying the thermalization \cite{Kidd2021,Brenes2021} 
and acts as a valuable detector of various phase transitions
\cite{Heyl2018,WangQ2019,Nie2020,ChenB2020,HuhK2021,Zamani2022,BinQ2023}.
At this point, we would like to mention that the properties of OTOC 
are obviously dependent on the kind of the observables. 
They are usually the physically relevant observables, such as the position and momentum operators, 
as well as spin and particle density operators in finite-range interacting systems.
But, the choice of them is usually motivated by certain specific physical problem.

In the present work, we employ the OTOC to analyze the dynamical signatures of 
the mixed eigenstates in the kicked top model. 
Before delving into this question and focusing on specific dynamical behaviors, let us
provide a brief review on the basic features of the kicked top model and the characterization 
of the mixed eigenstates.

\subsection{Kicked top model}

The kicked top model is a prototypical model in the studies of quantum chaos \cite{Haake2019} 
and can be experimentally realized in different platforms \cite{Chaudhury2009,Neill2016,Krithika2019,Meier2019}.
Its quantum version is described by the Hamiltonian 
\be \label{KTH}
   H=\alpha S_x+\frac{\kappa}{2S}S_z^2\sum_{n=-\infty}^{+\infty}\delta(t-n). 
\ee
Here, $S_\mu\ (\mu=x,y,z)$ are the angular momentum operators with total magnitude $S$ and satisfying
the standard commutation relations of angular momentum. 
The parameter $\alpha$ denotes the frequency of the free precession around $x$ axis, while $\kappa$
represents the strength of periodic $\delta$ kicks.
As the total angular momentum is a conserved quantity, 
the Hilbert space of $H$ (\ref{KTH}) has finite dimension equal to $2S+1$. 
Hence, we can investigate the dynamics of the model without the truncation of the Hilbert space.

It is known that the kick strength $\kappa$ controls the degree of chaos of the model, 
indicating a transition from integrability to chaos with increasing $\kappa$.
This is verified by the quasienergy statistics of the Floquet operator, which governs 
the time evolution between two successive kicks and is given by
\be \label{FloquetH}
   \mathcal{F}=\exp\left[-i\frac{\kappa}{2S}S_z^2\right]\exp(-i\alpha S_x).
\ee
The quasienergy spectrum of $H$ is then obtained through the eigenvalue equation
\be
  \mathcal{F}|\varepsilon_k\ra=e^{i\varepsilon_k}|\varepsilon_k\ra,
\ee
where $\varepsilon_k$ is the $k$th quasienergy associated to eigenstate $|\varepsilon_k\ra$.

The presence of chaos in the quantum kicked top model 
is a manifestation of integrability-to-chaos transition in its classical dynamics, 
which is given by \cite{Piga2019,Munoz2021,QwangR2023} 
\begin{align}\label{ClassicalDy}
&X_{m+1}=X_m\cos\Psi_m-(Y_m\cos\alpha-Z_m\sin\alpha)\sin\Psi_m, \notag \\
&Y_{m+1}=X_m\sin\Psi_m+(Y_m\cos\alpha+Z_m\sin\alpha)\cos\Psi_m, \notag \\
&Z_{m+1}=Y_m\sin\alpha+Z_m\cos\alpha,
\end{align}
where $\Psi_m=\kappa(Y_m\sin\alpha+Z_m\cos\alpha)$ and $\mathbf{X}=(X,Y,Z)=\la\mathbf{S}\ra/S$ 
are the classical dynamical variables.
The conservation of $\mathbf{S}^2$ implies $X^2+Y^2+Z^2=1$, 
which allows us to parameterize them
as $X=\sin\theta\cos\phi, Y=\sin\theta\sin\phi$, and $Z=\cos\theta$, 
with $\phi$ and $\theta$ being azimuthal and polar angles, respectively.
As a result, the classical phase space can be described by canonical variables 
$\phi=\arctan(Y/X)\in[-\pi,\pi]$ and $\cos\theta\in[-1,1]$.

 \begin{figure}
  \includegraphics[width=\textwidth]{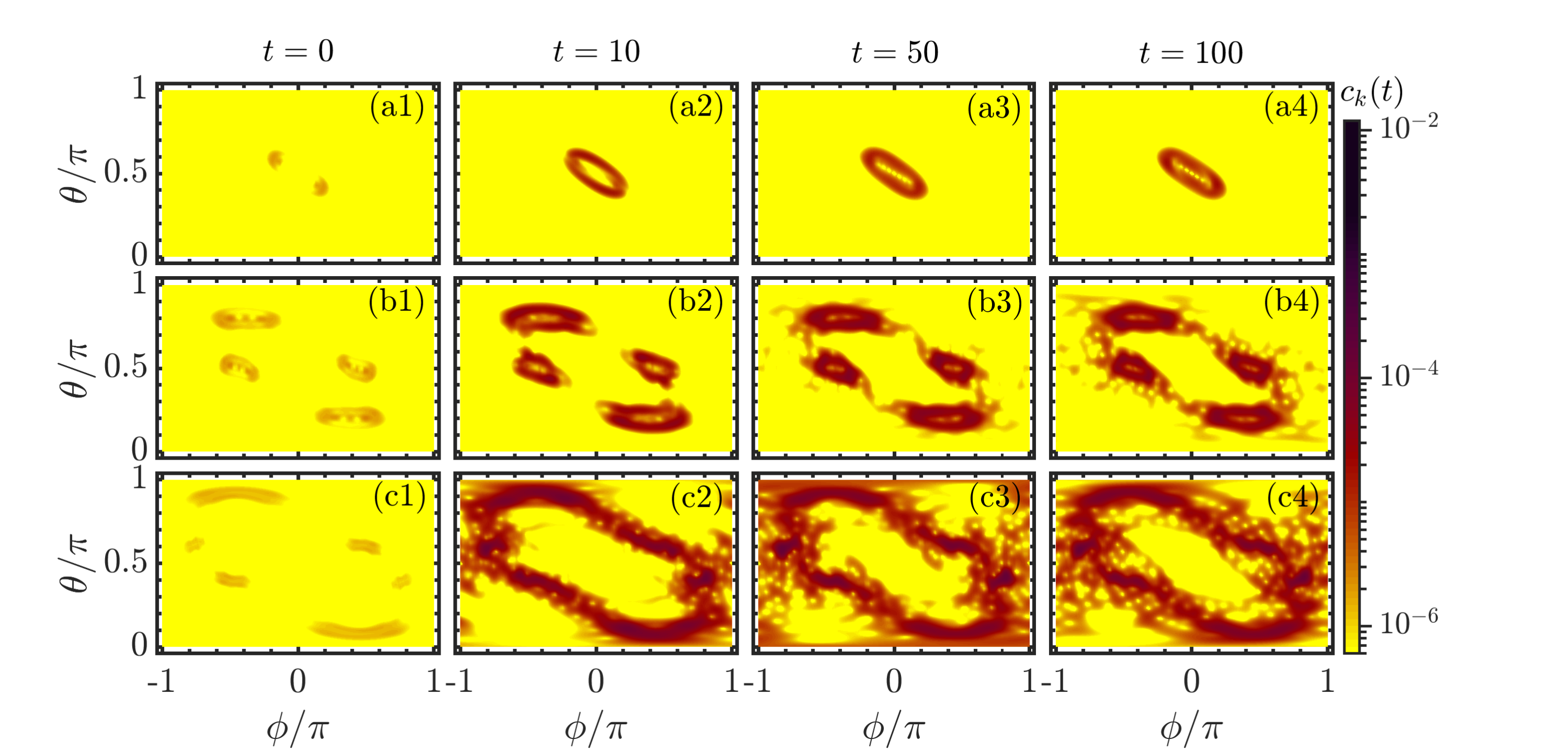}
  \caption{Snapshots of the OTOC $c_k(t)$, defined in Eq.~(\ref{OtcCh}), 
  at different time steps for the 
  eigenstates of the Floquet operator with $\chi_{6}=-1$ (a1)-(a4), 
  $\chi_{179}=0.3858$ (b1)-(b4), 
  and $\chi_{273}=0.9985$ (c1)-(c4). 
  Other parameters: $\alpha=13\pi/19$, $\kappa=3$, and the system size $S=150$.}
  \label{TotcCh}
 \end{figure}

A common way to show the transition to chaos in the 
classical dynamics is to examine the Poincar\'e section.  
It was known that the Poincar\'e section exhibits regular 
structure for integrable systems, defined by invariant tori, while it consists of
randomly scattered points for fully chaotic dynamics.
The Poincar\'e section obtained by solving Eq.~(\ref{ClassicalDy}) for different values of $\kappa$ 
with $\alpha=13\pi/19$ are plotted in Fig.~\ref{PSections}.
The variation of Poincar\'e section from regular pattern for small $\kappa$ 
to covered by randomly located points at large $\kappa$ is clearly 
indicating the transition to chaos with increasing $\kappa$. 

Our focus lies on the dynamical signatures of the mixed eigenstates, 
which exist in the quantum systems with mixed classical phase space 
consisting of regular islands embedded in the chaotic sea.
For our considered case, the mixed classical phase space is present for $2\lesssim\kappa\lesssim5.4$. 
We thus fixed $\kappa=3$ in our study.
The mixed feature of the classical phase space for $\kappa=3$ case can be visualized by 
the corresponding Poincar\'e section, as demonstrated in Fig.~\ref{PSections}(c). 
We have numerically verified that our main conclusions still hold 
for other values of $\kappa$, as long as the corresponding classical dynamics is mixed.
Moreover, a careful numerical check has shown that 
although the value of $\alpha$ can change the degree of chaos for both quantum and classical 
kicked top \cite{WangR2021}, it does not affect the main results of this work.
This allows us to fix $\alpha=13\pi/19$ in our study.

\subsection{Mixed eigenstates}

The mixed eigenstates, also referred to as hybridized states \cite{Varma2024}, are prevalent
in generic quantum systems that have mixed classical phase 
space with coexistence of regular and chaotic motions.
A prominent feature of the mixed eigenstates is manifested 
in their corresponding Husimi functions, 
distributing over both regular and chaotic regions.
This means that the types of eigenstates are encoded in their corresponding Husimi functions.

For the $k$th eigenstate, $|\varepsilon_k\ra$, of the Floquet operator, 
the Husimi function is given by \cite{Husimi1940}
\be \label{HsFD}
    Q_k(\theta,\phi)=|\la\theta,\phi|\varepsilon_k\ra|^2,
\ee 
where 
\be \label{CohS}
  |\theta,\phi\ra=e^{i\theta(S_x\sin\phi-S_y\cos\phi)}|S,S\ra 
\ee
are the $\mathrm{SU}(2)$ spin-coherent states \cite{Perelomov1977,ZhangW1990} 
localized at $(\theta,\phi)$ and $S_z|S,S\ra=S|S,S\ra$. 
Moreover, the Husimi function $Q_k(\theta,\phi)$ is normalized as
\be
   \frac{2S+1}{4\pi}\int d\mathcal{A}Q_k(\theta,\phi)=1,
\ee
with $d\mathcal{A}=\sin\theta d\theta d\phi$ being the area element on the unit sphere.
To identify the types of the eigenstates, let us first discretize the Husimi function
by dividing the classical phase space into a grid with $N\times N$ equal cells 
that are marked by their central points and are indexed as $(p,q)$ with $p,q=1,2,\ldots,N$. 
Then, we assign a value $C_{pq}=+1$ to the cells that reside in the chaotic region,
and $C_{pq}=-1$ otherwise.  
Finally, the types of the $k$th eigenstate is determined by the 
phase space overlap index, which is defined as
\be
   \chi_k=\frac{2S+1}{4\pi}\sum_{p,q}Q_k(\theta_p,\phi_q)C_{pq}\Delta A_{pq},
\ee 
where $\Delta A_{pq}$ is the area of the cell with index $(p,q)$.  
The definition of $\chi_k$ leads to $-1\leq\chi_k\leq1$ with $\chi_k=-1$ and $+1$, respectively, corresponding to
regular and fully chaotic eigenstates. 
Hence, the mixed eigenstates are identified as $\chi_k\neq\pm1$.

 \begin{figure}
  \includegraphics[width=\textwidth]{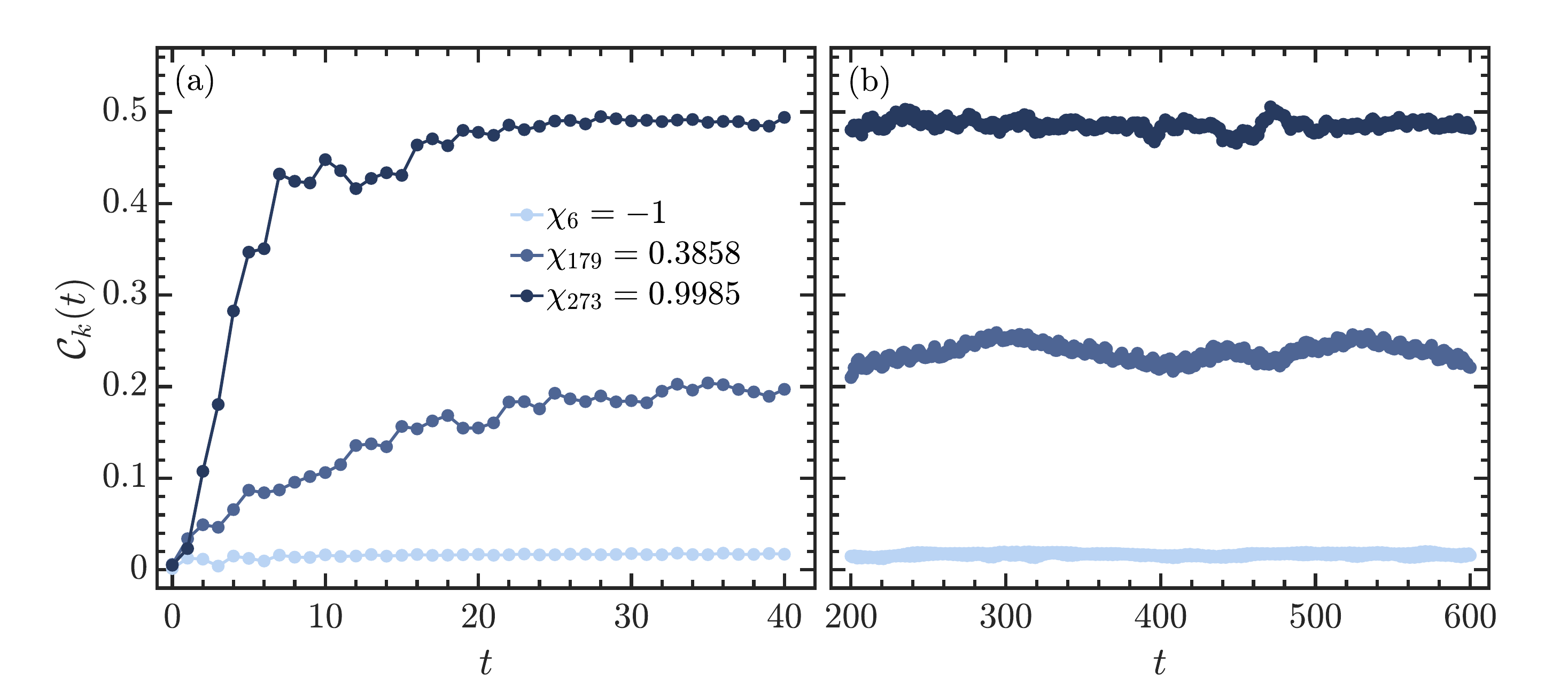}
  \caption{Short-time (a) and long-time (b) phase space averaged OTOC, $\mathcal{C}_k(t)$ in Eq.~(\ref{PSAvgOtc})
  for the eigenstates with corresponding values of $\chi_k$ are given in the legend of panel(a). 
  Other parameters: $\alpha=13\pi/19$, $\kappa=3$, and the system size $S=150$.}
  \label{AveragedCt}
 \end{figure}

In our numerical simulation, the chaotic region is generated by evolving an initial condition, 
which is randomly chosen from the chaotic region of phase space, up to $10^8$ kicks.
As a result, the complement includes all regular and possibly tiny chaotic regions.
However, we can take the tiny chaotic regions as a part of regular region, as
they are vanishingly small.    
Moreover, to ensure that the discretized Husimi function is 
normalized and our results are converged,
we set $N\times N=300\times300$ throughout this work.
The Husimi function for several quasieigenstates of $\mathcal{F}$ in (\ref{FloquetH})
with associated $\chi_k$ are shown in Fig.~\ref{Husimifs}. 
Comparing them to the corresponding Poincar\'e section, which we plot in Fig.~\ref{Husimifs}(e),   
we see that in contrast to the regular and chaotic eigenstates, 
such as the one displayed in Figs.~\ref{Husimifs}(a) and \ref{Husimifs}(d), 
the Husimi function of the mixed eigenstates spreads over both classical regular and chaotic regions. 
However, in the sufficiently deep semiclassical limit, the relative fraction 
of mixed-type eigenstates decays as a power law \cite{WangR2023,WangR2024,YanWR2024}.
This prominent character of the mixed eigenstates leads us to expect 
that it should also get reflected in the dynamical behaviors of mixed eigenstates.
In the following section, we investigate this question by means of the OTOC.

\section{Dynamical signatures of the mixed eigenstates} \label{Third}

To analyze how the mixed feature of the mixed eigenstates manifests 
in their dynamical behaviors through the OTOC, we take the initial Hermitian operators as
$A=s_x=S_x/S$ and $B=\rho_c=|\theta,\phi\ra\la\theta,\phi|$ 
with $|\theta,\phi\ra$ being the spin coherent state in (\ref{CohS}).
Then, for the $k$th eigenstate $|\varepsilon_k\ra$, the OTOC is given by
\be \label{OtcCh}
  c_k(t)=\la\varepsilon_k|[s_x(t),\rho_c]^\dag[s_x(t),\rho_c]|\varepsilon_k\ra
    =c_{k,1}(t)-c_{k,2}(t),
\ee 
where $c_{k,1}(t)=\la x(t)|x(t)\ra+\la y(t)|y(t)\ra$ and 
$c_{k,2}(t)=\la x(t)|y(t)\ra+\la y(t)|x(t)\ra$.
Here, we have defined $|x(t)\ra=s_x(t)\rho_c|\varepsilon_k\ra$ and $|y(t)\ra=\rho_cs_x(t)|\varepsilon_k\ra$
with $s_x(t)=\mathcal{F}^{\dag t}S_x\mathcal{F}^t/S$. 
The reason for choosing density operator of the coherent states is that 
it allows us to understand how the mixed structure in the phase space 
affects the evolution of the eigenstate OTOC. 
However, the choice of $s_x$ is not unique. 
In fact, we have checked that the results obtained for $s_x$ also hold for 
other operators, such as $s_y=S_y/S$, 
as long as they do not commut with the Floquet operator.
Otherwise, the OTOC would be independent of the time.

The role played by the type of the eigenstates for the behavior 
of $c_k(t)$ cannot be directly unveiled by Eq.~(\ref{OtcCh}). 
To uncover how the type of the eigenstates get reflected in the 
evolution of $c_k(t)$, we derive an upper bound of $c_k(t)$ in terms of 
the Husimi function of the $k$th eigenstate.
By using the closure relation of the eigenstates and spin coherent states,
\be \label{Identities}
   \sum_u|\varepsilon_u\ra\la\varepsilon_u|=1,\quad 
   \frac{2S+1}{4\pi}\int d\mathcal{A}|\theta,\phi\ra\la\theta,\phi|=1,
\ee
we first rewrite $c_k(t)$ as
\begin{align} 
  c_k(t)=\sum_u|\la\varepsilon_u|\mathcal{O}(t)|\varepsilon_k\ra|^2
    =\left(\frac{2S+1}{4\pi}\right)^2\sum_u\left|
      \int d\mathcal{A}_1\la\varepsilon_u|\mathcal{O}(t)|\theta_1,\phi_1\ra
      \la\theta_1,\phi_1|\varepsilon_k\ra\right|^2,
\end{align}
where $\mathcal{O}(t)=[s_x(t),\rho_c]$
is the commutator between $s_x(t)$ and $\rho_c$.
Then, the Cauchy-Schwarz inequality leads us to obtain
\begin{align}
   \left|\int d\mathcal{A}_1\la\varepsilon_u|\mathcal{O}(t)|\theta_1,\phi_1\ra
        \la\theta_1,\phi_1|\varepsilon_k\ra\right|^2
       &\leq\left(\int d\mathcal{A}_1 1^2\right)\left(\int d\mathcal{A}_1
       |\la\varepsilon_u|\mathcal{O}(t)|\theta_1,\phi_1\ra|^2Q_k(\theta_1,\phi_1)\right) \notag \\
       &=4\pi \int d\mathcal{A}_1
       Q_k(\theta_1,\phi_1)|\la\varepsilon_u|\mathcal{O}(t)|\theta_1,\phi_1\ra|^2,
\end{align} 
where $\int d\mathcal{A}_1=4\pi$ has been employed and 
$Q_k(\theta_1,\phi_1)$ denotes the Husimi function in Eq.~(\ref{HsFD}). 
As a result, we finally find that $c_k(t)$ is upper bounded by
\begin{align} \label{Rdfck}
  c_k(t)&\leq\frac{(2S+1)^2}{4\pi}\int d\mathcal{A}_1
     Q_k(\theta_1,\phi_1)\sum_u\la\theta_1,\phi_1|\mathcal{O}^\dag(t)|\varepsilon_u\ra
       \la\varepsilon_u|\mathcal{O}(t)|\theta_1,\phi_1\ra \notag \\
      &=\frac{(2S+1)^2}{4\pi}\int d\mathcal{A}_1Q_k(\theta_1,\phi_1)\mathcal{W}_t(\theta_1,\phi_1),
\end{align}
where $\mathcal{W}_t(\theta_1,\phi_1)=\la\theta_1,\phi_1|\mathcal{O}^\dag(t)
\mathcal{O}(t)|\theta_1,\phi_1\ra$ is the OTOC with respect to the spin coherent states.
Here, the last equality is obtained by using the closure 
relation of the eigenstates in (\ref{Identities}). 
   
It is known that the OTOC is less prone to scrambling for the 
coherent states located in the regular regions,
while it exhibits a fast spreading over the chaotic sea 
for the coherent states resided in the chaotic component \cite{Blocher2022,Varikuti2024}. 
Then, according to the inequality of (\ref{Rdfck}), $c_k(t)$ will evolve in the 
regular regions for the regular eigenstates and it should undergo 
a rapid expanding in the chaotic region for chaotic eigenstates.  
However, as the Husimi function of the mixed eigenstates occupies both regular and chaotic regions, 
the evolution of $c_k(t)$ will display some mixed features for the mixed eigenstates.
In other words, the time dependence of $c_k(t)$ for the mixed eigenstates is a combination of 
the restricting motion in the regular regions and an extending in the chaotic region.

In Fig.~\ref{TotcCh}, we plot $c_k(t)$ in classical phase space at 
different time steps for regular, mixed, and chaotic eigenstates, respectively.
The dependence of the behavior of $c_k(t)$ on the type of eigenstate is clearly visible.
Specifically, as seen in Figs.~\ref{TotcCh}(a1)-\ref{TotcCh}(a4), the evolution of OTOC for the 
regular state is confined within the regular region. 
On the contrary, the OTOC spreads over the chaotic region for the chaotic eigenstate, 
as demonstrated in Figs.~\ref{TotcCh}(c1)-\ref{TotcCh}(c4).
Distinct from both regular and chaotic eigenstates, 
as observed in Figs.~\ref{TotcCh}(b1)-\ref{TotcCh}(b4), 
the OTOC of the mixed eigenstate is initially evolved in 
the regular region and gradually spread into the chaotic region with increasing time.  
These results suggest that the type of the eigenstates leaves an imprint on the behavior of OTOC.

Further dynamical properties of the mixed eigenstates can be quantitatively revealed 
by the phase space averaged OTOC, which is defined as
\be \label{PSAvgOtc}
   \mathcal{C}_k(t)=\frac{2S+1}{4\pi}\int c_k(t)\sin\theta d\theta d\phi.
\ee 
The results in Fig.~\ref{TotcCh} indicate that
the time dependence of $\mathcal{C}_k(t)$ should be strongly correlated with the type of eigenstate.
This is verified in Fig.~\ref{AveragedCt}, where we show the short-time and long-time evolutions of
$\mathcal{C}_k(t)$ for the regular, mixed, and chaotic eigenstates.  
The confining behavior of $c_k(t)$ for the regular state results 
in $\mathcal{C}_k(t)$ evolving around a vanishingly small value with almost no fluctuations.  
In contrast, due to the fast spreading of $c_k(t)$ in the phase space, $C_k(t)$ of the chaotic state 
starts with a rapid growth followed by tiny fluctuations around certain saturation value.
In particular, one can observe that the behavior of $\mathcal{C}_k(t)$ for the 
mixed eigenstate is very different from the regular and chaotic states.
As the mixed eigenstate exhibits a slow extending of $c_k(t)$ over the phase space, 
the corresponding $\mathcal{C}_k(t)$ increases with a lower rate
and it eventually saturates with small oscillations at long time.
The saturation value of $\mathcal{C}_k(t)$ for the mixed eigenstate is 
smaller compared to the chaotic state and it should depend on the value of $\chi_k$. 
Moreover, we further note that the initial growth rate of $\mathcal{C}_k(t)$ also correlates to $\chi_k$.
These results imply that the dynamical signatures of the mixed eigenstates 
are encoded in both short- and long-time properties of $\mathcal{C}_k(t)$. 

\subsection{Short-time growth rate of phase space averaged OTOC}     
   
To analyze how the mixed eigenstates manifest themselves in
the short-time behavior of $\mathcal{C}_k(t)$, 
we define its initial growth rate as
\be
  \gamma_k=\frac{\mathcal{C}_k(\tau)-\mathcal{C}_k(0)}{\tau}
          \simeq\frac{d\mathcal{C}_k(\tau)}{d\tau},
\ee 
where $\tau$ is the finial time of the initial growth.
It was known that OTOCs usually exhibit a growth 
until the Ehrenfest time $t_E\sim\log(\mathcal{D}_\mathcal{H})/\lambda_{cl}$ 
with $\mathcal{D}_\mathcal{H}$ and $\lambda_{cl}$ are, respectively,
the Hilbert space dimension and the classical Lyapunov exponent.
As $\lambda_{cl}$ has the order of magnitude $O(1)$ for our considered case,
we thus take $\tau=\nint{\ln(2S+1)}$ in the numerical calculations.
Here, $\nint{x}$ means the nearest integer of $x$.

 \begin{figure}
  \includegraphics[width=\textwidth]{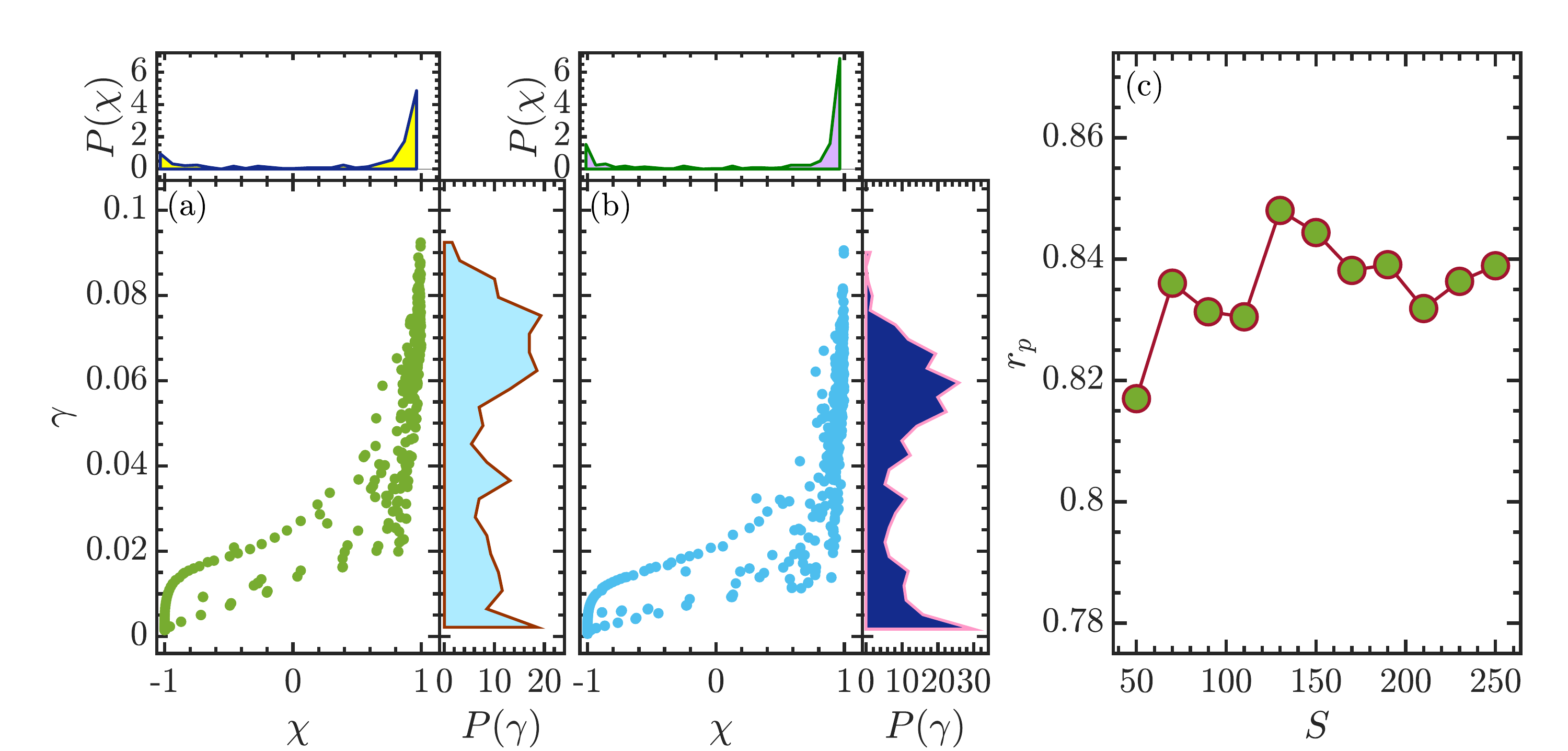}
  \caption{Scatter plots of $\gamma$ versus $\chi$ for the system size
  (a) $S=150$ and (b) $S=250$.
  The top and right panels in (a) and (b) are, respectively, plotted the 
  probability distributions $P(\chi)$ and $P(\gamma)$, defined in Eq.~(\ref{DefPdf}).
  (c) Pearson product-momentum correlation coefficient $r_p$ in Eq.~(\ref{PPMC}) 
  of $\chi_k$ and $\gamma_k$ as a function of system size $S$. 
  Other parameters: $\alpha=13\pi/19$, $\kappa=3$.}
  \label{Stpdf}
 \end{figure}

The results in Fig.~\ref{AveragedCt} indicate that
$\gamma_k$ is defined in an interval $\gamma_k\in[\gamma_{k,min}, \gamma_{k,max}]$.
For regular eigenstates, we have $\gamma_k=\gamma_{k,min}\sim0$, while
$\gamma_k=\gamma_{k,max}$ corresponding to the chaotic eigenstates.
The value of $\gamma_k$ for the mixed eigenstates varies in between. 

In Figs.~\ref{Stpdf}(a) and \ref{Stpdf}(b), we show the scatter plots 
of $\gamma$ and $\chi$ of the eigenstates for different system sizes.
As expected, we see that the regular and chaotic eigenstates have the values of 
$\gamma$ that are clustered around its minimal and maximal value, 
regardless of the system size. 
For the mixed eigenstates with $-1<\chi<1$, 
the initial growth rate $\gamma$ exhibits a wide 
distribution between two extreme values, indicating that 
the value of $\gamma$ can be used to characterize the degree of 
mixture of the mixed eigenstates. 
 
To further show how the mixed eigenstates are correlated with $\gamma$, we 
consider the probability distributions of $\gamma$ and $\chi$,
which are defined as
\be \label{DefPdf}
   P(\gamma)=\frac{1}{2S+1}\sum_{\gamma_k}\delta(\gamma-\gamma_k),\quad
   P(\chi)=\frac{1}{2S+1}\sum_{\chi_k}\delta(\chi-\chi_k),
\ee
where $S$ is the magnitude of the angular momentum and also denotes the system size.
Our previous works \cite{WangR2023,WangR2024} have demonstrated that $P(\chi)$ has the double peak shape
with two peaks corresponding to regular and chaotic eigenstates, respectively. 
Consequently, one can expect that the shape of $P(\gamma)$ should also be the double peak.
This is verified in top and right panels of Fig.~\ref{Stpdf}(a) and \ref{Stpdf}(b), 
where the probability distribution $P(\gamma)$ and $P(\chi)$ are displayed.
However, we note that $P(\gamma)$ is supported on a much smaller interval than $P(\chi)$, which leads to
a less sharp double peak in $P(\gamma)$ as compared to $P(\chi)$.
Nevertheless, we observe that the sharpness of the double peak shape in $P(\chi)$ 
can be enhanced as the system size is increased. 
It is worth pointing out that this enhancement
also implies the decreasing of the relative fraction of the mixed eigenstates 
with increasing the system size,
consisting with our previous result \cite{WangR2023,YanWR2024}.

The quality of the correlation can be quantitatively measured by the correlation coefficient.
For the two random variables $\{(U_i,V_i)\}$, the dependence between them  
is usually quantified by the well-known Pearson product-moment correlation coefficient \cite{Sneyd2022}, 
defined by
\be \label{PPMC}
    r_p=\frac{\sum_i(U_i-\bar{U})(V_i-\bar{V})}{\sqrt{\sum_i(U_i-\bar{U})^2\sum_i(V_i-\bar{V})^2}},
\ee
where $\bar{U}$ and $\bar{V}$ are the average of $U_i$ and $V_i$, respectively. 
The Pearson coefficient $r_P$ for our case is calculated by replacing $(U_i,V_i)$ with $(\chi_k,\gamma_k)$. 
In Fig.~\ref{Stpdf}(c), we plot how $r_P$ varies with increasing the system size $S$.  
It is obvious that $r_P$ has a larger value $r_P\approx 0.84$
almost independent of the system size. 
This feature confirms the strong correlation between $\gamma$ and $\chi$. 
It also prompts us to conjecture that the short time growth rate of 
the phase space averaged OTOC reflects 
the dynamical character of the mixed eigenstates.

 \begin{figure}
  \includegraphics[width=\textwidth]{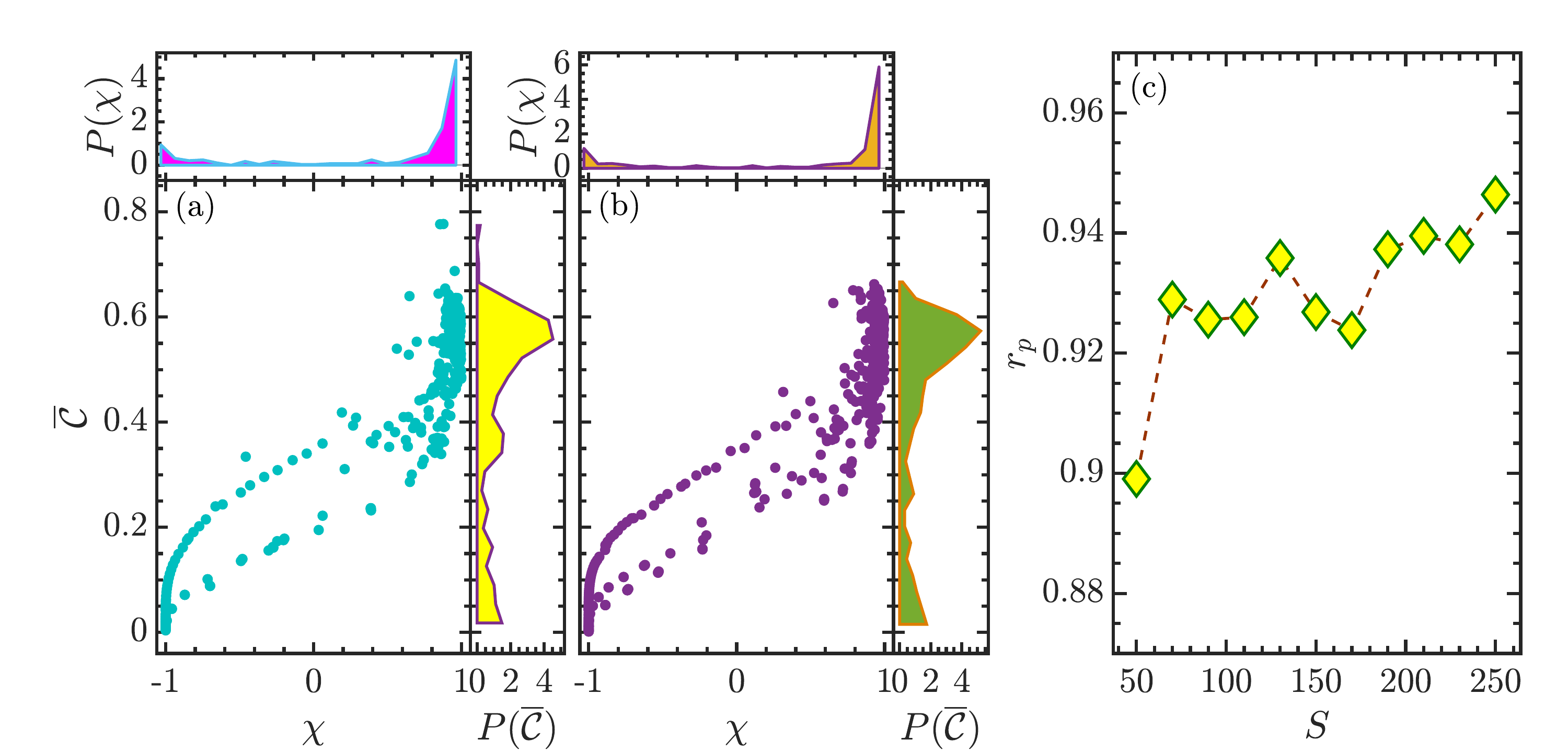}
  \caption{Scatter plots of $\overline{\mathcal{C}}$ as a function of 
  $\chi$ for the system size (a) $S=150$ and (b) $S=250$.
  The top and right panels in (a) and (b) are, respectively, plotted the 
  probability distributions $P(\chi)$ in Eq.~(\ref{DefPdf}) and $P(\overline{\mathcal{C}})$ in Eq.~(\ref{PdfTav}).
  (c) Pearson product-momentum correlation coefficient $r_p$, defined in Eq.~(\ref{PPMC}), 
  of $\overline{\mathcal{C}}_k$ and $\chi_k$ as a function of system size $S$. 
  Other parameters: $\alpha=13\pi/19$, $\kappa=3$.}
  \label{Avcpdf}
 \end{figure}

\subsection{Long-time average of phase space averaged OTOC}

The dynamical signature of the mixed eigenstates can also be revealed by the long-time 
average of $\mathcal{C}_k(t)$, defined as 
 \be \label{TavgOtc}
   \overline{\mathcal{C}}_k=\lim_{T\to\infty}\frac{1}{T}\int_t^{t+T}\mathcal{C}_k(v)dv,
 \ee
 where $t\gg1$ should be sufficiently larger than the initial time scale.
 In our numerical simulations, we take $t=100$ and $T=500$.
 We have carefully checked that further increasing $t$ and $T$ does not change our main results.
 By inserting Eq.~(\ref{PSAvgOtc}) into (\ref{TavgOtc}), after some algebra, one gets
 \be \label{Tavgct}
   \overline{\mathcal{C}}_k=\frac{2S+1}{4\pi}\int\bar{c}_k\sin\theta d\theta d\phi,
 \ee
 where
 \be \label{Avct}
   \bar{c}_k=\lim_{T\to\infty}\frac{1}{T}\int_t^{t+T}c_k(v) dv=C_{k,1}-2C_{k,2},
 \ee
 and 
 \begin{align}
    C_{k,1}&=Q_k\sum_pQ_p(s^2_x)_{pp}+\sum_pQ_p(s_x)_{kp}(s_x)_{pk}-2(s_x)^2_{kk}Q^4_k, \notag \\
    C_{k,2}&=Q_k\sum_{p\neq k}\left[Q_p(s_x)_{pp}(s_x)_{kk}+Q_p(s_x)_{kp}(s_x)_{pk}\right],
 \end{align} 
 with $Q_k$ being the Husimi function (\ref{HsFD}) of the $k$th eigenstate and
 $(s_x)_{ab}=\la\varepsilon_a|S_x|\varepsilon_b\ra/S$.
 Here, the integration in Eq.~(\ref{Avct}) has been carried out by assuming that the  
 energy spectrum has no degeneracies.
 
It is known that the fully chaotic eigenstates are almost uniformly distributed over a given basis.
As a result, we have $C_{k,1}\sim O(1/S)$ and $C_{k,2}\sim O(1/S^2)$, 
suggesting $\bar{c}_k\sim O(1/S)$ and $\overline{\mathcal{C}}_{k,max}\sim O(1)$.  
For the regular eigenstates, $C_{k,1}$ and $C_{k,2}$ have the same oder of magnitude $O(1/S^2)$,
leading to $\overline{\mathcal{C}}_{k,min}\sim O(1/S)$. 
However, as the mixed eigenstates are partially localized 
in the regular region and partially extended in the chaotic region, 
one can reasonably expect that the values of their $\overline{\mathcal{C}}_k$
are directly linked to $\chi_k$ and vary from
$\overline{\mathcal{C}}_{k,min}$ to $\overline{\mathcal{C}}_{k,max}$.

The scatter plots of $\overline{\mathcal{C}}_k$ versus $\chi$ for all eigenstates with different system sizes are
shown in Figs.~\ref{Avcpdf}(a) and \ref{Avcpdf}(b). 
An overall similarity between Figs.~\ref{Avcpdf}(a)-\ref{Avcpdf}(b) 
and Figs.~\ref{Stpdf}(a)-\ref{Stpdf}(b) is clearly visible.   
We see that $\overline{\mathcal{C}}$ clusters around its minimal and maximal values
which are, respectively, corresponding to the regular and chaotic eigenstates, 
as indicated by the value of $\chi$. 
Besides, one can further observe that the mixed eigenstates 
with $-1<\chi<1$ have values of $\overline{\mathcal{C}}_k$  
that are scattered between two clusters. 
These results confirm that the types of the eigenstates have strong impact on 
evolution of the OTOC, and the unique features in the OTOC exhibited by
the mixed eigenstates enable us to distinguish them 
from chaotic and regular eigenstates. 

To further uncover the links between the long-time averaged OTOC and 
the signatures of mixed eigenstates, we consider the 
probability distribution of $\overline{\mathcal{C}}_k$, defined by
\be \label{PdfTav}
   P(\overline{\mathcal{C}})=\frac{1}{2S+1}\sum_k\delta(\overline{\mathcal{C}}-\overline{\mathcal{C}}_k),
\ee
and compare it to $P(\chi)$ in Eq.~(\ref{DefPdf}).
The results for different systems are plotted in the right 
and top panels of Figs.~\ref{Avcpdf}(a) and \ref{Avcpdf}(b). 
We first note that $P(\overline{\mathcal{C}})$ has a similar shape as $P(\chi)$,
regardless of the system size.
Both $P(\overline{\mathcal{C}})$ and $P(\chi)$ are characterized 
by double peak distribution with two peaks corresponding to 
regular and chaotic eigenstates, respectively. 
The mixed eigenstates are marked by smaller values of 
$P(\overline{\mathcal{C}})$ and $P(\chi)$ that are distributed between two peaks. 
In particular, we observe that the double peak shape of $P(\overline{\mathcal{C}})$ 
and $P(\chi)$ become sharper as the system size is increased.
This is in agreement with the expectation that the relative fraction of the mixed eigenstates 
decreases with increasing the system size 
\cite{Lozej2022,WangR2023,WangR2024,YanR2024,YanWR2024}.    

 The remarkable agreements exhibited by $P(\overline{\mathcal{C}})$ 
 and $P(\chi)$ indicate the equivalence between them.
 To confirm this statement, we study the correlation coefficient between 
 $\overline{\mathcal{C}}_k$ and $\chi_k$, claculated by $r_p$ in 
 Eq.~(\ref{PPMC}) with $(\overline{\mathcal{C}}_k,\chi_k)$ serving as $(U_i,V_i)$.
 In Fig.~\ref{Avcpdf}(c), we plot how $r_p$ varies as a function of the system size $S$.
 We see that $r_p$ shows a weak dependence on the system size and it displays only 
 a small fluctuation around $r_p\approx 0.93$ with increasing $S$. 
 This not only demonstrates that the behavior of the long-time averaged OTOC is 
 strongly correlated with the types of the eigenstates, but also
 verifies that the OTOC acts as a valuable tool to analyze 
 the dynamical signatures of the mixed eigenstates.
 Moreover, one can observe that $r_p$ in Fig.~\ref{Avcpdf}(c) is larger 
 than the one in Fig.~\ref{Stpdf}(c), suggesting that 
 the long-time averaged OTOC is more reliable to distinguish the 
 mixed eigenstates than the initial growth rate of the OTOC.

\section{Conclusions} \label{Fourth}

In conclusion, we have examined the dynamical signatures 
of the mixed eigenstates in a mixed-type system.
Different from regular and fully chaotic systems,
a mixed-type system has mixed 
phase space with regular islands embedded in chaotic sea. 
This led Percival to classify quantum spectra of the mixed-type systems 
into regular and chaotic types \cite{Percival1973}.
However, this binary classification is an idealization and 
the quantum eigenstates in the actual situations are more complex.
In fact, it has been found that the mixed eigenstates with Husimi function 
occupying both regular and chaotic regions are more prevalent in the mixed-type systems. 
Hence, studying the properties of the mixed eigenstates is crucial 
for understanding various phenomena exhibited by the mixed-type systems.
Previous works have investigated the statistical 
features of the mixed eigenstates \cite{Varma2024} and had 
numerically verified that their relative fraction in the 
semiclassical limit decreases according to a power law 
\cite{Lozej2022,WangR2023,WangR2024,YanR2024,YanWR2024}.  
In this work, we have demonstrated how to characterize the 
mixed eigenstates through the dynamics of the OTOC
in the kicked top model, which behaves as a mixed-type system for certain control parameters.

The mixed eigenstates in our study are identified using the phase space
overlap index, which measures the degree of the mixture of the eigenstates
and has been employed in our previous works 
\cite{Lozej2022,WangR2023,WangR2024,YanR2024,YanWR2024}.   
We have shown that the time dependence of OTOC for the mixed eigenstates exhibits unique 
behavior which distinguishes the cases of regular and chaotic eigenstates.
In particular, the short- and long-time behaviors of OTOC exhibit an obvious dependence  
on the degree of the mixture of the eigenstates.
This has led us to analyze the dynamical characters of the mixed eigenstates
through the initial growth rate and long-time average of the OTOC, respectively.
We have revealed how the initial growth rate and long-time average of OTOC 
connect to the phase space overlap index and quantified their 
correlations via the Pearson product-moment correlation coefficient.

Our findings offer further insights into the features of mixed eigenstates and 
also provide a comprehensive perspective on the mixed-type systems.
As the mixed eigenstates are commonly described by the distribution 
of their Husimi function in both regular and chaotic regions, we
expect that the main conclusions of this work may still hold for other mixed-type systems.
It would be interesting to systematically study dynamical properties of the mixed eigenstates
in various mixed-type systems, such as billiards and Dicke model, via the OTOCs. 
Another question that deserves future exploring is 
to find an analytical explanation for our numerical results.
Moreover, the OTOCs have been experimentally measured in a variety 
of platforms \cite{LiJun2017,Garttner2017,Braumuller2022,Green2022}. 
This led us to further expect that our work could stimulate more experimental studies 
of dynamics in mixed-type systems.

\acknowledgments  

This work was supported by the Slovenian Research and Innovation Agency (ARIS) under the 
Grants Nos.~J1-4387 and P1-0306.

\bibliographystyle{apsrev4-1}


\begin{thebibliography}{105}%
\makeatletter
\providecommand \@ifxundefined [1]{%
 \@ifx{#1\undefined}
}%
\providecommand \@ifnum [1]{%
 \ifnum #1\expandafter \@firstoftwo
 \else \expandafter \@secondoftwo
 \fi
}%
\providecommand \@ifx [1]{%
 \ifx #1\expandafter \@firstoftwo
 \else \expandafter \@secondoftwo
 \fi
}%
\providecommand \natexlab [1]{#1}%
\providecommand \enquote  [1]{``#1''}%
\providecommand \bibnamefont  [1]{#1}%
\providecommand \bibfnamefont [1]{#1}%
\providecommand \citenamefont [1]{#1}%
\providecommand \href@noop [0]{\@secondoftwo}%
\providecommand \href [0]{\begingroup \@sanitize@url \@href}%
\providecommand \@href[1]{\@@startlink{#1}\@@href}%
\providecommand \@@href[1]{\endgroup#1\@@endlink}%
\providecommand \@sanitize@url [0]{\catcode `\\12\catcode `\$12\catcode
  `\&12\catcode `\#12\catcode `\^12\catcode `\_12\catcode `\%12\relax}%
\providecommand \@@startlink[1]{}%
\providecommand \@@endlink[0]{}%
\providecommand \url  [0]{\begingroup\@sanitize@url \@url }%
\providecommand \@url [1]{\endgroup\@href {#1}{\urlprefix }}%
\providecommand \urlprefix  [0]{URL }%
\providecommand \Eprint [0]{\href }%
\providecommand \doibase [0]{http://dx.doi.org/}%
\providecommand \selectlanguage [0]{\@gobble}%
\providecommand \bibinfo  [0]{\@secondoftwo}%
\providecommand \bibfield  [0]{\@secondoftwo}%
\providecommand \translation [1]{[#1]}%
\providecommand \BibitemOpen [0]{}%
\providecommand \bibitemStop [0]{}%
\providecommand \bibitemNoStop [0]{.\EOS\space}%
\providecommand \EOS [0]{\spacefactor3000\relax}%
\providecommand \BibitemShut  [1]{\csname bibitem#1\endcsname}%
\let\auto@bib@innerbib\@empty
\bibitem [{\citenamefont {Das}\ and\ \citenamefont {Ghosh}(2023)}]{Das2023}%
  \BibitemOpen
  \bibfield  {author} {\bibinfo {author} {\bibfnamefont {A.~K.}\ \bibnamefont
  {Das}}\ and\ \bibinfo {author} {\bibfnamefont {A.}~\bibnamefont {Ghosh}},\
  }\href {\doibase 10.1088/1751-8121/ad0b5a} {\bibfield  {journal} {\bibinfo
  {journal} {Journal of Physics A: Mathematical and Theoretical}\ }\textbf
  {\bibinfo {volume} {56}},\ \bibinfo {pages} {495003} (\bibinfo {year}
  {2023})}\BibitemShut {NoStop}%
\bibitem [{\citenamefont {Shi}\ \emph {et~al.}(2023)\citenamefont {Shi},
  \citenamefont {Vardhan},\ and\ \citenamefont {Liu}}]{ShiZ2023}%
  \BibitemOpen
  \bibfield  {author} {\bibinfo {author} {\bibfnamefont {Z.~D.}\ \bibnamefont
  {Shi}}, \bibinfo {author} {\bibfnamefont {S.}~\bibnamefont {Vardhan}}, \ and\
  \bibinfo {author} {\bibfnamefont {H.}~\bibnamefont {Liu}},\ }\href {\doibase
  10.1103/PhysRevB.108.224305} {\bibfield  {journal} {\bibinfo  {journal}
  {Phys. Rev. B}\ }\textbf {\bibinfo {volume} {108}},\ \bibinfo {pages}
  {224305} (\bibinfo {year} {2023})}\BibitemShut {NoStop}%
\bibitem [{\citenamefont {Luca~D'Alessio}\ and\ \citenamefont
  {Rigol}(2016)}]{Luca2016}%
  \BibitemOpen
  \bibfield  {author} {\bibinfo {author} {\bibfnamefont {A.~P.}\ \bibnamefont
  {Luca~D'Alessio}, \bibfnamefont {Yariv~Kafri}}\ and\ \bibinfo {author}
  {\bibfnamefont {M.}~\bibnamefont {Rigol}},\ }\href {\doibase
  10.1080/00018732.2016.1198134} {\bibfield  {journal} {\bibinfo  {journal}
  {Advances in Physics}\ }\textbf {\bibinfo {volume} {65}},\ \bibinfo {pages}
  {239} (\bibinfo {year} {2016})}\BibitemShut {NoStop}%
\bibitem [{\citenamefont {Torres-Herrera}\ and\ \citenamefont
  {Santos}(2019)}]{Herrera2019}%
  \BibitemOpen
  \bibfield  {author} {\bibinfo {author} {\bibfnamefont {E.~J.}\ \bibnamefont
  {Torres-Herrera}}\ and\ \bibinfo {author} {\bibfnamefont {L.~F.}\
  \bibnamefont {Santos}},\ }\href {\doibase 10.1140/epjst/e2019-800057-8}
  {\bibfield  {journal} {\bibinfo  {journal} {The European Physical Journal
  Special Topics}\ }\textbf {\bibinfo {volume} {227}},\ \bibinfo {pages} {1897}
  (\bibinfo {year} {2019})}\BibitemShut {NoStop}%
\bibitem [{\citenamefont {Pandey}\ \emph {et~al.}(2020)\citenamefont {Pandey},
  \citenamefont {Claeys}, \citenamefont {Campbell}, \citenamefont
  {Polkovnikov},\ and\ \citenamefont {Sels}}]{Pandey2020}%
  \BibitemOpen
  \bibfield  {author} {\bibinfo {author} {\bibfnamefont {M.}~\bibnamefont
  {Pandey}}, \bibinfo {author} {\bibfnamefont {P.~W.}\ \bibnamefont {Claeys}},
  \bibinfo {author} {\bibfnamefont {D.~K.}\ \bibnamefont {Campbell}}, \bibinfo
  {author} {\bibfnamefont {A.}~\bibnamefont {Polkovnikov}}, \ and\ \bibinfo
  {author} {\bibfnamefont {D.}~\bibnamefont {Sels}},\ }\href {\doibase
  10.1103/PhysRevX.10.041017} {\bibfield  {journal} {\bibinfo  {journal} {Phys.
  Rev. X}\ }\textbf {\bibinfo {volume} {10}},\ \bibinfo {pages} {041017}
  (\bibinfo {year} {2020})}\BibitemShut {NoStop}%
\bibitem [{\citenamefont {Nandkishore}\ and\ \citenamefont
  {Huse}(2015)}]{Nandkishore2015}%
  \BibitemOpen
  \bibfield  {author} {\bibinfo {author} {\bibfnamefont {R.}~\bibnamefont
  {Nandkishore}}\ and\ \bibinfo {author} {\bibfnamefont {D.~A.}\ \bibnamefont
  {Huse}},\ }\href {\doibase
  https://doi.org/10.1146/annurev-conmatphys-031214-014726} {\bibfield
  {journal} {\bibinfo  {journal} {Annual Review of Condensed Matter Physics}\
  }\textbf {\bibinfo {volume} {6}},\ \bibinfo {pages} {15} (\bibinfo {year}
  {2015})}\BibitemShut {NoStop}%
\bibitem [{\citenamefont {Abanin}\ \emph {et~al.}(2019)\citenamefont {Abanin},
  \citenamefont {Altman}, \citenamefont {Bloch},\ and\ \citenamefont
  {Serbyn}}]{Abanin2019}%
  \BibitemOpen
  \bibfield  {author} {\bibinfo {author} {\bibfnamefont {D.~A.}\ \bibnamefont
  {Abanin}}, \bibinfo {author} {\bibfnamefont {E.}~\bibnamefont {Altman}},
  \bibinfo {author} {\bibfnamefont {I.}~\bibnamefont {Bloch}}, \ and\ \bibinfo
  {author} {\bibfnamefont {M.}~\bibnamefont {Serbyn}},\ }\href {\doibase
  10.1103/RevModPhys.91.021001} {\bibfield  {journal} {\bibinfo  {journal}
  {Rev. Mod. Phys.}\ }\textbf {\bibinfo {volume} {91}},\ \bibinfo {pages}
  {021001} (\bibinfo {year} {2019})}\BibitemShut {NoStop}%
\bibitem [{\citenamefont {Georgescu}\ \emph {et~al.}(2014)\citenamefont
  {Georgescu}, \citenamefont {Ashhab},\ and\ \citenamefont
  {Nori}}]{Georgescu2014}%
  \BibitemOpen
  \bibfield  {author} {\bibinfo {author} {\bibfnamefont {I.~M.}\ \bibnamefont
  {Georgescu}}, \bibinfo {author} {\bibfnamefont {S.}~\bibnamefont {Ashhab}}, \
  and\ \bibinfo {author} {\bibfnamefont {F.}~\bibnamefont {Nori}},\ }\href
  {\doibase 10.1103/RevModPhys.86.153} {\bibfield  {journal} {\bibinfo
  {journal} {Rev. Mod. Phys.}\ }\textbf {\bibinfo {volume} {86}},\ \bibinfo
  {pages} {153} (\bibinfo {year} {2014})}\BibitemShut {NoStop}%
\bibitem [{\citenamefont {Daley}\ \emph {et~al.}(2022)\citenamefont {Daley},
  \citenamefont {Bloch}, \citenamefont {Kokail}, \citenamefont {Flannigan},
  \citenamefont {Pearson}, \citenamefont {Troyer},\ and\ \citenamefont
  {Zoller}}]{Daley2022}%
  \BibitemOpen
  \bibfield  {author} {\bibinfo {author} {\bibfnamefont {A.~J.}\ \bibnamefont
  {Daley}}, \bibinfo {author} {\bibfnamefont {I.}~\bibnamefont {Bloch}},
  \bibinfo {author} {\bibfnamefont {C.}~\bibnamefont {Kokail}}, \bibinfo
  {author} {\bibfnamefont {S.}~\bibnamefont {Flannigan}}, \bibinfo {author}
  {\bibfnamefont {N.}~\bibnamefont {Pearson}}, \bibinfo {author} {\bibfnamefont
  {M.}~\bibnamefont {Troyer}}, \ and\ \bibinfo {author} {\bibfnamefont
  {P.}~\bibnamefont {Zoller}},\ }\href {\doibase 10.1038/s41586-022-04940-6}
  {\bibfield  {journal} {\bibinfo  {journal} {Nature}\ }\textbf {\bibinfo
  {volume} {607}},\ \bibinfo {pages} {667} (\bibinfo {year}
  {2022})}\BibitemShut {NoStop}%
\bibitem [{\citenamefont {Giovannetti}\ \emph {et~al.}(2011)\citenamefont
  {Giovannetti}, \citenamefont {Lloyd},\ and\ \citenamefont
  {Maccone}}]{Giovannetti2011}%
  \BibitemOpen
  \bibfield  {author} {\bibinfo {author} {\bibfnamefont {V.}~\bibnamefont
  {Giovannetti}}, \bibinfo {author} {\bibfnamefont {S.}~\bibnamefont {Lloyd}},
  \ and\ \bibinfo {author} {\bibfnamefont {L.}~\bibnamefont {Maccone}},\ }\href
  {\doibase 10.1038/nphoton.2011.35} {\bibfield  {journal} {\bibinfo  {journal}
  {Nature Photonics}\ }\textbf {\bibinfo {volume} {5}},\ \bibinfo {pages} {222}
  (\bibinfo {year} {2011})}\BibitemShut {NoStop}%
\bibitem [{\citenamefont {Magdalena~Szczykulska}\ and\ \citenamefont
  {Datta}(2016)}]{Magdalena2016}%
  \BibitemOpen
  \bibfield  {author} {\bibinfo {author} {\bibfnamefont {T.~B.}\ \bibnamefont
  {Magdalena~Szczykulska}}\ and\ \bibinfo {author} {\bibfnamefont
  {A.}~\bibnamefont {Datta}},\ }\href {\doibase 10.1080/23746149.2016.1230476}
  {\bibfield  {journal} {\bibinfo  {journal} {Advances in Physics: X}\ }\textbf
  {\bibinfo {volume} {1}},\ \bibinfo {pages} {621} (\bibinfo {year}
  {2016})}\BibitemShut {NoStop}%
\bibitem [{\citenamefont {Percival}(1973)}]{Percival1973}%
  \BibitemOpen
  \bibfield  {author} {\bibinfo {author} {\bibfnamefont {I.~C.}\ \bibnamefont
  {Percival}},\ }\href {\doibase 10.1088/0022-3700/6/9/002} {\bibfield
  {journal} {\bibinfo  {journal} {Journal of Physics B: Atomic and Molecular
  Physics}\ }\textbf {\bibinfo {volume} {6}},\ \bibinfo {pages} {L229}
  (\bibinfo {year} {1973})}\BibitemShut {NoStop}%
\bibitem [{\citenamefont {Stechel}\ and\ \citenamefont
  {Heller}(1984)}]{Stechel1984}%
  \BibitemOpen
  \bibfield  {author} {\bibinfo {author} {\bibfnamefont {E.~B.}\ \bibnamefont
  {Stechel}}\ and\ \bibinfo {author} {\bibfnamefont {E.~J.}\ \bibnamefont
  {Heller}},\ }\href {\doibase
  https://doi.org/10.1146/annurev.pc.35.100184.003023} {\bibfield  {journal}
  {\bibinfo  {journal} {Annual Review of Physical Chemistry}\ }\textbf
  {\bibinfo {volume} {35}},\ \bibinfo {pages} {563} (\bibinfo {year}
  {1984})}\BibitemShut {NoStop}%
\bibitem [{\citenamefont {Berry}(1977)}]{Berry1977}%
  \BibitemOpen
  \bibfield  {author} {\bibinfo {author} {\bibfnamefont {M.~V.}\ \bibnamefont
  {Berry}},\ }\href {\doibase 10.1088/0305-4470/10/12/016} {\bibfield
  {journal} {\bibinfo  {journal} {Journal of Physics A: Mathematical and
  General}\ }\textbf {\bibinfo {volume} {10}},\ \bibinfo {pages} {2083}
  (\bibinfo {year} {1977})}\BibitemShut {NoStop}%
\bibitem [{\citenamefont {Berry}\ and\ \citenamefont
  {Ziman}(1977)}]{Victor1977}%
  \BibitemOpen
  \bibfield  {author} {\bibinfo {author} {\bibfnamefont {M.~V.}\ \bibnamefont
  {Berry}}\ and\ \bibinfo {author} {\bibfnamefont {J.~M.}\ \bibnamefont
  {Ziman}},\ }\href {\doibase 10.1098/rsta.1977.0145} {\bibfield  {journal}
  {\bibinfo  {journal} {Philosophical Transactions of the Royal Society of
  London. Series A, Mathematical and Physical Sciences}\ }\textbf {\bibinfo
  {volume} {287}},\ \bibinfo {pages} {237} (\bibinfo {year}
  {1977})}\BibitemShut {NoStop}%
\bibitem [{\citenamefont {Robnik}(2000)}]{Robnik2000}%
  \BibitemOpen
  \bibfield  {author} {\bibinfo {author} {\bibfnamefont {M.}~\bibnamefont
  {Robnik}},\ }\href {https://arxiv.org/abs/nlin/0003058} {\enquote {\bibinfo
  {title} {Topics in quantum chaos of generic systems},}\ } (\bibinfo {year}
  {2000}),\ \Eprint {http://arxiv.org/abs/nlin/0003058} {arXiv:nlin/0003058
  [nlin.CD]} \BibitemShut {NoStop}%
\bibitem [{\citenamefont {Berry}\ and\ \citenamefont
  {Robnik}(1984)}]{BerryR1984}%
  \BibitemOpen
  \bibfield  {author} {\bibinfo {author} {\bibfnamefont {M.~V.}\ \bibnamefont
  {Berry}}\ and\ \bibinfo {author} {\bibfnamefont {M.}~\bibnamefont {Robnik}},\
  }\href {\doibase 10.1088/0305-4470/17/12/013} {\bibfield  {journal} {\bibinfo
   {journal} {Journal of Physics A: Mathematical and General}\ }\textbf
  {\bibinfo {volume} {17}},\ \bibinfo {pages} {2413} (\bibinfo {year}
  {1984})}\BibitemShut {NoStop}%
\bibitem [{\citenamefont {Robnik}(2020)}]{Robnik2020}%
  \BibitemOpen
  \bibfield  {author} {\bibinfo {author} {\bibfnamefont {M.}~\bibnamefont
  {Robnik}},\ }\href@noop {} {\bibfield  {journal} {\bibinfo  {journal}
  {Nonlinear Phenomena in Complex Systems}\ }\textbf {\bibinfo {volume} {23}},\
  \bibinfo {pages} {172} (\bibinfo {year} {2020})}\BibitemShut {NoStop}%
\bibitem [{\citenamefont {Robnik}(2024)}]{Robnik2024}%
  \BibitemOpen
  \bibfield  {author} {\bibinfo {author} {\bibfnamefont {M.}~\bibnamefont
  {Robnik}},\ }\href {\doibase 10.5281/zenodo.10030989} {\bibfield  {journal}
  {\bibinfo  {journal} {{Interdisciplinary Journal Nonlinear Phenomena in
  Complex Systems}}\ }\textbf {\bibinfo {volume} {26}} (\bibinfo {year}
  {2024}),\ 10.5281/zenodo.10030989}\BibitemShut {NoStop}%
\bibitem [{\citenamefont {Tomsovic}\ and\ \citenamefont
  {Ullmo}(1994)}]{Tomsovic1994}%
  \BibitemOpen
  \bibfield  {author} {\bibinfo {author} {\bibfnamefont {S.}~\bibnamefont
  {Tomsovic}}\ and\ \bibinfo {author} {\bibfnamefont {D.}~\bibnamefont
  {Ullmo}},\ }\href {\doibase 10.1103/PhysRevE.50.145} {\bibfield  {journal}
  {\bibinfo  {journal} {Phys. Rev. E}\ }\textbf {\bibinfo {volume} {50}},\
  \bibinfo {pages} {145} (\bibinfo {year} {1994})}\BibitemShut {NoStop}%
\bibitem [{\citenamefont {Frischat}\ and\ \citenamefont
  {Doron}(1998)}]{Frischat1998}%
  \BibitemOpen
  \bibfield  {author} {\bibinfo {author} {\bibfnamefont {S.~D.}\ \bibnamefont
  {Frischat}}\ and\ \bibinfo {author} {\bibfnamefont {E.}~\bibnamefont
  {Doron}},\ }\href {\doibase 10.1103/PhysRevE.57.1421} {\bibfield  {journal}
  {\bibinfo  {journal} {Phys. Rev. E}\ }\textbf {\bibinfo {volume} {57}},\
  \bibinfo {pages} {1421} (\bibinfo {year} {1998})}\BibitemShut {NoStop}%
\bibitem [{\citenamefont {L\"ock}\ \emph {et~al.}(2010)\citenamefont {L\"ock},
  \citenamefont {B\"acker}, \citenamefont {Ketzmerick},\ and\ \citenamefont
  {Schlagheck}}]{Lock2010}%
  \BibitemOpen
  \bibfield  {author} {\bibinfo {author} {\bibfnamefont {S.}~\bibnamefont
  {L\"ock}}, \bibinfo {author} {\bibfnamefont {A.}~\bibnamefont {B\"acker}},
  \bibinfo {author} {\bibfnamefont {R.}~\bibnamefont {Ketzmerick}}, \ and\
  \bibinfo {author} {\bibfnamefont {P.}~\bibnamefont {Schlagheck}},\ }\href
  {\doibase 10.1103/PhysRevLett.104.114101} {\bibfield  {journal} {\bibinfo
  {journal} {Phys. Rev. Lett.}\ }\textbf {\bibinfo {volume} {104}},\ \bibinfo
  {pages} {114101} (\bibinfo {year} {2010})}\BibitemShut {NoStop}%
\bibitem [{\citenamefont {Lozej}\ \emph {et~al.}(2022)\citenamefont {Lozej},
  \citenamefont {Lukman},\ and\ \citenamefont {Robnik}}]{Lozej2022}%
  \BibitemOpen
  \bibfield  {author} {\bibinfo {author} {\bibfnamefont {{\ifmmode
  \check{C}\else \v{C}\fi{}}.}~\bibnamefont {Lozej}}, \bibinfo {author}
  {\bibfnamefont {D.}~\bibnamefont {Lukman}}, \ and\ \bibinfo {author}
  {\bibfnamefont {M.}~\bibnamefont {Robnik}},\ }\href {\doibase
  10.1103/PhysRevE.106.054203} {\bibfield  {journal} {\bibinfo  {journal}
  {Phys. Rev. E}\ }\textbf {\bibinfo {volume} {106}},\ \bibinfo {pages}
  {054203} (\bibinfo {year} {2022})}\BibitemShut {NoStop}%
\bibitem [{\citenamefont {Wang}\ and\ \citenamefont
  {Robnik}(2023{\natexlab{a}})}]{WangR2023}%
  \BibitemOpen
  \bibfield  {author} {\bibinfo {author} {\bibfnamefont {Q.}~\bibnamefont
  {Wang}}\ and\ \bibinfo {author} {\bibfnamefont {M.}~\bibnamefont {Robnik}},\
  }\href {\doibase 10.1103/PhysRevE.108.054217} {\bibfield  {journal} {\bibinfo
   {journal} {Phys. Rev. E}\ }\textbf {\bibinfo {volume} {108}},\ \bibinfo
  {pages} {054217} (\bibinfo {year} {2023}{\natexlab{a}})}\BibitemShut
  {NoStop}%
\bibitem [{\citenamefont {Wang}\ and\ \citenamefont
  {Robnik}(2024)}]{WangR2024}%
  \BibitemOpen
  \bibfield  {author} {\bibinfo {author} {\bibfnamefont {Q.}~\bibnamefont
  {Wang}}\ and\ \bibinfo {author} {\bibfnamefont {M.}~\bibnamefont {Robnik}},\
  }\href {\doibase 10.1103/PhysRevE.109.024225} {\bibfield  {journal} {\bibinfo
   {journal} {Phys. Rev. E}\ }\textbf {\bibinfo {volume} {109}},\ \bibinfo
  {pages} {024225} (\bibinfo {year} {2024})}\BibitemShut {NoStop}%
\bibitem [{\citenamefont {Varma}\ \emph {et~al.}(2024)\citenamefont {Varma},
  \citenamefont {Vardi},\ and\ \citenamefont {Cohen}}]{Varma2024}%
  \BibitemOpen
  \bibfield  {author} {\bibinfo {author} {\bibfnamefont {A.~V.}\ \bibnamefont
  {Varma}}, \bibinfo {author} {\bibfnamefont {A.}~\bibnamefont {Vardi}}, \ and\
  \bibinfo {author} {\bibfnamefont {D.}~\bibnamefont {Cohen}},\ }\href
  {\doibase 10.1103/PhysRevE.109.064207} {\bibfield  {journal} {\bibinfo
  {journal} {Phys. Rev. E}\ }\textbf {\bibinfo {volume} {109}},\ \bibinfo
  {pages} {064207} (\bibinfo {year} {2024})}\BibitemShut {NoStop}%
\bibitem [{\citenamefont {Yan}\ and\ \citenamefont {Robnik}(2024)}]{YanR2024}%
  \BibitemOpen
  \bibfield  {author} {\bibinfo {author} {\bibfnamefont {H.}~\bibnamefont
  {Yan}}\ and\ \bibinfo {author} {\bibfnamefont {M.}~\bibnamefont {Robnik}},\
  }\href {\doibase 10.1103/PhysRevE.109.054211} {\bibfield  {journal} {\bibinfo
   {journal} {Phys. Rev. E}\ }\textbf {\bibinfo {volume} {109}},\ \bibinfo
  {pages} {054211} (\bibinfo {year} {2024})}\BibitemShut {NoStop}%
\bibitem [{\citenamefont {Yan}\ \emph {et~al.}(2024)\citenamefont {Yan},
  \citenamefont {Wang},\ and\ \citenamefont {Robnik}}]{YanWR2024}%
  \BibitemOpen
  \bibfield  {author} {\bibinfo {author} {\bibfnamefont {H.}~\bibnamefont
  {Yan}}, \bibinfo {author} {\bibfnamefont {Q.}~\bibnamefont {Wang}}, \ and\
  \bibinfo {author} {\bibfnamefont {M.}~\bibnamefont {Robnik}},\ }\href
  {\doibase 10.1103/PhysRevE.110.064222} {\bibfield  {journal} {\bibinfo
  {journal} {Phys. Rev. E}\ }\textbf {\bibinfo {volume} {110}},\ \bibinfo
  {pages} {064222} (\bibinfo {year} {2024})}\BibitemShut {NoStop}%
\bibitem [{\citenamefont {Xu}\ and\ \citenamefont
  {Swingle}(2019)}]{XuSheng2019}%
  \BibitemOpen
  \bibfield  {author} {\bibinfo {author} {\bibfnamefont {S.}~\bibnamefont
  {Xu}}\ and\ \bibinfo {author} {\bibfnamefont {B.}~\bibnamefont {Swingle}},\
  }\href {\doibase 10.1103/PhysRevX.9.031048} {\bibfield  {journal} {\bibinfo
  {journal} {Phys. Rev. X}\ }\textbf {\bibinfo {volume} {9}},\ \bibinfo {pages}
  {031048} (\bibinfo {year} {2019})}\BibitemShut {NoStop}%
\bibitem [{\citenamefont {Gonz\'alez~Alonso}\ \emph {et~al.}(2019)\citenamefont
  {Gonz\'alez~Alonso}, \citenamefont {Yunger~Halpern},\ and\ \citenamefont
  {Dressel}}]{Gonzalez2019}%
  \BibitemOpen
  \bibfield  {author} {\bibinfo {author} {\bibfnamefont {J.~R.}\ \bibnamefont
  {Gonz\'alez~Alonso}}, \bibinfo {author} {\bibfnamefont {N.}~\bibnamefont
  {Yunger~Halpern}}, \ and\ \bibinfo {author} {\bibfnamefont {J.}~\bibnamefont
  {Dressel}},\ }\href {\doibase 10.1103/PhysRevLett.122.040404} {\bibfield
  {journal} {\bibinfo  {journal} {Phys. Rev. Lett.}\ }\textbf {\bibinfo
  {volume} {122}},\ \bibinfo {pages} {040404} (\bibinfo {year}
  {2019})}\BibitemShut {NoStop}%
\bibitem [{\citenamefont {Yan}\ \emph {et~al.}(2020)\citenamefont {Yan},
  \citenamefont {Cincio},\ and\ \citenamefont {Zurek}}]{YanB2020}%
  \BibitemOpen
  \bibfield  {author} {\bibinfo {author} {\bibfnamefont {B.}~\bibnamefont
  {Yan}}, \bibinfo {author} {\bibfnamefont {L.}~\bibnamefont {Cincio}}, \ and\
  \bibinfo {author} {\bibfnamefont {W.~H.}\ \bibnamefont {Zurek}},\ }\href
  {\doibase 10.1103/PhysRevLett.124.160603} {\bibfield  {journal} {\bibinfo
  {journal} {Phys. Rev. Lett.}\ }\textbf {\bibinfo {volume} {124}},\ \bibinfo
  {pages} {160603} (\bibinfo {year} {2020})}\BibitemShut {NoStop}%
\bibitem [{\citenamefont {Xu}\ and\ \citenamefont
  {Swingle}(2024)}]{XuSheng2024}%
  \BibitemOpen
  \bibfield  {author} {\bibinfo {author} {\bibfnamefont {S.}~\bibnamefont
  {Xu}}\ and\ \bibinfo {author} {\bibfnamefont {B.}~\bibnamefont {Swingle}},\
  }\href {\doibase 10.1103/PRXQuantum.5.010201} {\bibfield  {journal} {\bibinfo
   {journal} {PRX Quantum}\ }\textbf {\bibinfo {volume} {5}},\ \bibinfo {pages}
  {010201} (\bibinfo {year} {2024})}\BibitemShut {NoStop}%
\bibitem [{\citenamefont {Larkin}\ and\ \citenamefont
  {Ovchinnikov}(1969)}]{Larkin1969}%
  \BibitemOpen
  \bibfield  {author} {\bibinfo {author} {\bibfnamefont {A.~I.}\ \bibnamefont
  {Larkin}}\ and\ \bibinfo {author} {\bibfnamefont {Y.~N.}\ \bibnamefont
  {Ovchinnikov}},\ }\href@noop {} {\bibfield  {journal} {\bibinfo  {journal}
  {Sov Phys JETP}\ }\textbf {\bibinfo {volume} {28}},\ \bibinfo {pages} {1200}
  (\bibinfo {year} {1969})}\BibitemShut {NoStop}%
\bibitem [{\citenamefont {Chen}\ \emph {et~al.}(2017)\citenamefont {Chen},
  \citenamefont {Zhou}, \citenamefont {Huse},\ and\ \citenamefont
  {Fradkin}}]{ChenX2017}%
  \BibitemOpen
  \bibfield  {author} {\bibinfo {author} {\bibfnamefont {X.}~\bibnamefont
  {Chen}}, \bibinfo {author} {\bibfnamefont {T.}~\bibnamefont {Zhou}}, \bibinfo
  {author} {\bibfnamefont {D.~A.}\ \bibnamefont {Huse}}, \ and\ \bibinfo
  {author} {\bibfnamefont {E.}~\bibnamefont {Fradkin}},\ }\href {\doibase
  https://doi.org/10.1002/andp.201600332} {\bibfield  {journal} {\bibinfo
  {journal} {Annalen der Physik}\ }\textbf {\bibinfo {volume} {529}},\ \bibinfo
  {pages} {1600332} (\bibinfo {year} {2017})}\BibitemShut {NoStop}%
\bibitem [{\citenamefont {Fan}\ \emph {et~al.}(2017)\citenamefont {Fan},
  \citenamefont {Zhang}, \citenamefont {Shen},\ and\ \citenamefont
  {Zhai}}]{RFan2017}%
  \BibitemOpen
  \bibfield  {author} {\bibinfo {author} {\bibfnamefont {R.}~\bibnamefont
  {Fan}}, \bibinfo {author} {\bibfnamefont {P.}~\bibnamefont {Zhang}}, \bibinfo
  {author} {\bibfnamefont {H.}~\bibnamefont {Shen}}, \ and\ \bibinfo {author}
  {\bibfnamefont {H.}~\bibnamefont {Zhai}},\ }\href {\doibase
  https://doi.org/10.1016/j.scib.2017.04.011} {\bibfield  {journal} {\bibinfo
  {journal} {Science Bulletin}\ }\textbf {\bibinfo {volume} {62}},\ \bibinfo
  {pages} {707} (\bibinfo {year} {2017})}\BibitemShut {NoStop}%
\bibitem [{\citenamefont {D\'ora}\ and\ \citenamefont
  {Moessner}(2017)}]{Dora2017}%
  \BibitemOpen
  \bibfield  {author} {\bibinfo {author} {\bibfnamefont {B.}~\bibnamefont
  {D\'ora}}\ and\ \bibinfo {author} {\bibfnamefont {R.}~\bibnamefont
  {Moessner}},\ }\href {\doibase 10.1103/PhysRevLett.119.026802} {\bibfield
  {journal} {\bibinfo  {journal} {Phys. Rev. Lett.}\ }\textbf {\bibinfo
  {volume} {119}},\ \bibinfo {pages} {026802} (\bibinfo {year}
  {2017})}\BibitemShut {NoStop}%
\bibitem [{\citenamefont {Lin}\ and\ \citenamefont
  {Motrunich}(2018)}]{LinC2018}%
  \BibitemOpen
  \bibfield  {author} {\bibinfo {author} {\bibfnamefont {C.-J.}\ \bibnamefont
  {Lin}}\ and\ \bibinfo {author} {\bibfnamefont {O.~I.}\ \bibnamefont
  {Motrunich}},\ }\href {\doibase 10.1103/PhysRevB.97.144304} {\bibfield
  {journal} {\bibinfo  {journal} {Phys. Rev. B}\ }\textbf {\bibinfo {volume}
  {97}},\ \bibinfo {pages} {144304} (\bibinfo {year} {2018})}\BibitemShut
  {NoStop}%
\bibitem [{\citenamefont {Shenker}\ and\ \citenamefont
  {Stanford}(2014)}]{Shenker2014}%
  \BibitemOpen
  \bibfield  {author} {\bibinfo {author} {\bibfnamefont {S.~H.}\ \bibnamefont
  {Shenker}}\ and\ \bibinfo {author} {\bibfnamefont {D.}~\bibnamefont
  {Stanford}},\ }\href {\doibase 10.1007/JHEP03(2014)067} {\bibfield  {journal}
  {\bibinfo  {journal} {Journal of High Energy Physics}\ }\textbf {\bibinfo
  {volume} {2014}},\ \bibinfo {pages} {67} (\bibinfo {year}
  {2014})}\BibitemShut {NoStop}%
\bibitem [{\citenamefont {Maldacena}\ \emph {et~al.}(2016)\citenamefont
  {Maldacena}, \citenamefont {Shenker},\ and\ \citenamefont
  {Stanford}}]{Maldacena2016}%
  \BibitemOpen
  \bibfield  {author} {\bibinfo {author} {\bibfnamefont {J.}~\bibnamefont
  {Maldacena}}, \bibinfo {author} {\bibfnamefont {S.~H.}\ \bibnamefont
  {Shenker}}, \ and\ \bibinfo {author} {\bibfnamefont {D.}~\bibnamefont
  {Stanford}},\ }\href {\doibase 10.1007/JHEP08(2016)106} {\bibfield  {journal}
  {\bibinfo  {journal} {Journal of High Energy Physics}\ }\textbf {\bibinfo
  {volume} {2016}},\ \bibinfo {pages} {106} (\bibinfo {year}
  {2016})}\BibitemShut {NoStop}%
\bibitem [{\citenamefont {Rozenbaum}\ \emph {et~al.}(2017)\citenamefont
  {Rozenbaum}, \citenamefont {Ganeshan},\ and\ \citenamefont
  {Galitski}}]{Rozenbaum2017}%
  \BibitemOpen
  \bibfield  {author} {\bibinfo {author} {\bibfnamefont {E.~B.}\ \bibnamefont
  {Rozenbaum}}, \bibinfo {author} {\bibfnamefont {S.}~\bibnamefont {Ganeshan}},
  \ and\ \bibinfo {author} {\bibfnamefont {V.}~\bibnamefont {Galitski}},\
  }\href {\doibase 10.1103/PhysRevLett.118.086801} {\bibfield  {journal}
  {\bibinfo  {journal} {Phys. Rev. Lett.}\ }\textbf {\bibinfo {volume} {118}},\
  \bibinfo {pages} {086801} (\bibinfo {year} {2017})}\BibitemShut {NoStop}%
\bibitem [{\citenamefont {Garc\'{\i}a-Mata}\ \emph {et~al.}(2018)\citenamefont
  {Garc\'{\i}a-Mata}, \citenamefont {Saraceno}, \citenamefont {Jalabert},
  \citenamefont {Roncaglia},\ and\ \citenamefont {Wisniacki}}]{Garcia2018}%
  \BibitemOpen
  \bibfield  {author} {\bibinfo {author} {\bibfnamefont {I.}~\bibnamefont
  {Garc\'{\i}a-Mata}}, \bibinfo {author} {\bibfnamefont {M.}~\bibnamefont
  {Saraceno}}, \bibinfo {author} {\bibfnamefont {R.~A.}\ \bibnamefont
  {Jalabert}}, \bibinfo {author} {\bibfnamefont {A.~J.}\ \bibnamefont
  {Roncaglia}}, \ and\ \bibinfo {author} {\bibfnamefont {D.~A.}\ \bibnamefont
  {Wisniacki}},\ }\href {\doibase 10.1103/PhysRevLett.121.210601} {\bibfield
  {journal} {\bibinfo  {journal} {Phys. Rev. Lett.}\ }\textbf {\bibinfo
  {volume} {121}},\ \bibinfo {pages} {210601} (\bibinfo {year}
  {2018})}\BibitemShut {NoStop}%
\bibitem [{\citenamefont {Ch\'avez-Carlos}\ \emph {et~al.}(2019)\citenamefont
  {Ch\'avez-Carlos}, \citenamefont {L\'opez-del Carpio}, \citenamefont
  {Bastarrachea-Magnani}, \citenamefont {Str\'ansk\'y}, \citenamefont
  {Lerma-Hern\'andez}, \citenamefont {Santos},\ and\ \citenamefont
  {Hirsch}}]{Carlos2019}%
  \BibitemOpen
  \bibfield  {author} {\bibinfo {author} {\bibfnamefont {J.}~\bibnamefont
  {Ch\'avez-Carlos}}, \bibinfo {author} {\bibfnamefont {B.}~\bibnamefont
  {L\'opez-del Carpio}}, \bibinfo {author} {\bibfnamefont {M.~A.}\ \bibnamefont
  {Bastarrachea-Magnani}}, \bibinfo {author} {\bibfnamefont {P.}~\bibnamefont
  {Str\'ansk\'y}}, \bibinfo {author} {\bibfnamefont {S.}~\bibnamefont
  {Lerma-Hern\'andez}}, \bibinfo {author} {\bibfnamefont {L.~F.}\ \bibnamefont
  {Santos}}, \ and\ \bibinfo {author} {\bibfnamefont {J.~G.}\ \bibnamefont
  {Hirsch}},\ }\href {\doibase 10.1103/PhysRevLett.122.024101} {\bibfield
  {journal} {\bibinfo  {journal} {Phys. Rev. Lett.}\ }\textbf {\bibinfo
  {volume} {122}},\ \bibinfo {pages} {024101} (\bibinfo {year}
  {2019})}\BibitemShut {NoStop}%
\bibitem [{\citenamefont {Lewis-Swan}\ \emph {et~al.}(2019)\citenamefont
  {Lewis-Swan}, \citenamefont {Safavi-Naini}, \citenamefont {Bollinger},\ and\
  \citenamefont {Rey}}]{Lewis2019}%
  \BibitemOpen
  \bibfield  {author} {\bibinfo {author} {\bibfnamefont {R.~J.}\ \bibnamefont
  {Lewis-Swan}}, \bibinfo {author} {\bibfnamefont {A.}~\bibnamefont
  {Safavi-Naini}}, \bibinfo {author} {\bibfnamefont {J.~J.}\ \bibnamefont
  {Bollinger}}, \ and\ \bibinfo {author} {\bibfnamefont {A.~M.}\ \bibnamefont
  {Rey}},\ }\href {\doibase 10.1038/s41467-019-09436-y} {\bibfield  {journal}
  {\bibinfo  {journal} {Nature Communications}\ }\textbf {\bibinfo {volume}
  {10}},\ \bibinfo {pages} {1581} (\bibinfo {year} {2019})}\BibitemShut
  {NoStop}%
\bibitem [{\citenamefont {Aleiner}\ \emph {et~al.}(2016)\citenamefont
  {Aleiner}, \citenamefont {Faoro},\ and\ \citenamefont {Ioffe}}]{Aleiner2016}%
  \BibitemOpen
  \bibfield  {author} {\bibinfo {author} {\bibfnamefont {I.~L.}\ \bibnamefont
  {Aleiner}}, \bibinfo {author} {\bibfnamefont {L.}~\bibnamefont {Faoro}}, \
  and\ \bibinfo {author} {\bibfnamefont {L.~B.}\ \bibnamefont {Ioffe}},\ }\href
  {\doibase https://doi.org/10.1016/j.aop.2016.09.006} {\bibfield  {journal}
  {\bibinfo  {journal} {Annals of Physics}\ }\textbf {\bibinfo {volume}
  {375}},\ \bibinfo {pages} {378} (\bibinfo {year} {2016})}\BibitemShut
  {NoStop}%
\bibitem [{\citenamefont {Cotler}\ \emph {et~al.}(2018)\citenamefont {Cotler},
  \citenamefont {Ding},\ and\ \citenamefont {Penington}}]{Cotler2018}%
  \BibitemOpen
  \bibfield  {author} {\bibinfo {author} {\bibfnamefont {J.~S.}\ \bibnamefont
  {Cotler}}, \bibinfo {author} {\bibfnamefont {D.}~\bibnamefont {Ding}}, \ and\
  \bibinfo {author} {\bibfnamefont {G.~R.}\ \bibnamefont {Penington}},\ }\href
  {\doibase https://doi.org/10.1016/j.aop.2018.07.020} {\bibfield  {journal}
  {\bibinfo  {journal} {Annals of Physics}\ }\textbf {\bibinfo {volume}
  {396}},\ \bibinfo {pages} {318} (\bibinfo {year} {2018})}\BibitemShut
  {NoStop}%
\bibitem [{\citenamefont {Fortes}\ \emph {et~al.}(2019)\citenamefont {Fortes},
  \citenamefont {Garc\'{\i}a-Mata}, \citenamefont {Jalabert},\ and\
  \citenamefont {Wisniacki}}]{Fortes2019}%
  \BibitemOpen
  \bibfield  {author} {\bibinfo {author} {\bibfnamefont {E.~M.}\ \bibnamefont
  {Fortes}}, \bibinfo {author} {\bibfnamefont {I.}~\bibnamefont
  {Garc\'{\i}a-Mata}}, \bibinfo {author} {\bibfnamefont {R.~A.}\ \bibnamefont
  {Jalabert}}, \ and\ \bibinfo {author} {\bibfnamefont {D.~A.}\ \bibnamefont
  {Wisniacki}},\ }\href {\doibase 10.1103/PhysRevE.100.042201} {\bibfield
  {journal} {\bibinfo  {journal} {Phys. Rev. E}\ }\textbf {\bibinfo {volume}
  {100}},\ \bibinfo {pages} {042201} (\bibinfo {year} {2019})}\BibitemShut
  {NoStop}%
\bibitem [{\citenamefont {Xu}\ \emph {et~al.}(2020)\citenamefont {Xu},
  \citenamefont {Scaffidi},\ and\ \citenamefont {Cao}}]{XuT2020}%
  \BibitemOpen
  \bibfield  {author} {\bibinfo {author} {\bibfnamefont {T.}~\bibnamefont
  {Xu}}, \bibinfo {author} {\bibfnamefont {T.}~\bibnamefont {Scaffidi}}, \ and\
  \bibinfo {author} {\bibfnamefont {X.}~\bibnamefont {Cao}},\ }\href {\doibase
  10.1103/PhysRevLett.124.140602} {\bibfield  {journal} {\bibinfo  {journal}
  {Phys. Rev. Lett.}\ }\textbf {\bibinfo {volume} {124}},\ \bibinfo {pages}
  {140602} (\bibinfo {year} {2020})}\BibitemShut {NoStop}%
\bibitem [{\citenamefont {Kirkby}\ \emph {et~al.}(2021)\citenamefont {Kirkby},
  \citenamefont {O'Dell},\ and\ \citenamefont {Mumford}}]{Kirkby2021}%
  \BibitemOpen
  \bibfield  {author} {\bibinfo {author} {\bibfnamefont {W.}~\bibnamefont
  {Kirkby}}, \bibinfo {author} {\bibfnamefont {D.~H.~J.}\ \bibnamefont
  {O'Dell}}, \ and\ \bibinfo {author} {\bibfnamefont {J.}~\bibnamefont
  {Mumford}},\ }\href {\doibase 10.1103/PhysRevA.104.043308} {\bibfield
  {journal} {\bibinfo  {journal} {Phys. Rev. A}\ }\textbf {\bibinfo {volume}
  {104}},\ \bibinfo {pages} {043308} (\bibinfo {year} {2021})}\BibitemShut
  {NoStop}%
\bibitem [{\citenamefont {Dowling}\ \emph {et~al.}(2023)\citenamefont
  {Dowling}, \citenamefont {Kos},\ and\ \citenamefont {Modi}}]{Dowling2023}%
  \BibitemOpen
  \bibfield  {author} {\bibinfo {author} {\bibfnamefont {N.}~\bibnamefont
  {Dowling}}, \bibinfo {author} {\bibfnamefont {P.}~\bibnamefont {Kos}}, \ and\
  \bibinfo {author} {\bibfnamefont {K.}~\bibnamefont {Modi}},\ }\href {\doibase
  10.1103/PhysRevLett.131.180403} {\bibfield  {journal} {\bibinfo  {journal}
  {Phys. Rev. Lett.}\ }\textbf {\bibinfo {volume} {131}},\ \bibinfo {pages}
  {180403} (\bibinfo {year} {2023})}\BibitemShut {NoStop}%
\bibitem [{\citenamefont {Rozenbaum}\ \emph {et~al.}(2019)\citenamefont
  {Rozenbaum}, \citenamefont {Ganeshan},\ and\ \citenamefont
  {Galitski}}]{Rozenbaum2019}%
  \BibitemOpen
  \bibfield  {author} {\bibinfo {author} {\bibfnamefont {E.~B.}\ \bibnamefont
  {Rozenbaum}}, \bibinfo {author} {\bibfnamefont {S.}~\bibnamefont {Ganeshan}},
  \ and\ \bibinfo {author} {\bibfnamefont {V.}~\bibnamefont {Galitski}},\
  }\href {\doibase 10.1103/PhysRevB.100.035112} {\bibfield  {journal} {\bibinfo
   {journal} {Phys. Rev. B}\ }\textbf {\bibinfo {volume} {100}},\ \bibinfo
  {pages} {035112} (\bibinfo {year} {2019})}\BibitemShut {NoStop}%
\bibitem [{\citenamefont {Rautenberg}\ and\ \citenamefont
  {G\"arttner}(2020)}]{Rautenberg2020}%
  \BibitemOpen
  \bibfield  {author} {\bibinfo {author} {\bibfnamefont {M.}~\bibnamefont
  {Rautenberg}}\ and\ \bibinfo {author} {\bibfnamefont {M.}~\bibnamefont
  {G\"arttner}},\ }\href {\doibase 10.1103/PhysRevA.101.053604} {\bibfield
  {journal} {\bibinfo  {journal} {Phys. Rev. A}\ }\textbf {\bibinfo {volume}
  {101}},\ \bibinfo {pages} {053604} (\bibinfo {year} {2020})}\BibitemShut
  {NoStop}%
\bibitem [{\citenamefont {Alonso}\ \emph {et~al.}(2022)\citenamefont {Alonso},
  \citenamefont {Shammah}, \citenamefont {Ahmed}, \citenamefont {Nori},\ and\
  \citenamefont {Dressel}}]{Alonso2022}%
  \BibitemOpen
  \bibfield  {author} {\bibinfo {author} {\bibfnamefont {J.~R.~G.}\
  \bibnamefont {Alonso}}, \bibinfo {author} {\bibfnamefont {N.}~\bibnamefont
  {Shammah}}, \bibinfo {author} {\bibfnamefont {S.}~\bibnamefont {Ahmed}},
  \bibinfo {author} {\bibfnamefont {F.}~\bibnamefont {Nori}}, \ and\ \bibinfo
  {author} {\bibfnamefont {J.}~\bibnamefont {Dressel}},\ }\href
  {https://arxiv.org/abs/2201.08175} {\enquote {\bibinfo {title} {Diagnosing
  quantum chaos with out-of-time-ordered-correlator quasiprobability in the
  kicked-top model},}\ } (\bibinfo {year} {2022}),\ \Eprint
  {http://arxiv.org/abs/2201.08175} {arXiv:2201.08175 [quant-ph]} \BibitemShut
  {NoStop}%
\bibitem [{\citenamefont {Trunin}(2023)}]{Trunin2023}%
  \BibitemOpen
  \bibfield  {author} {\bibinfo {author} {\bibfnamefont {D.~A.}\ \bibnamefont
  {Trunin}},\ }\href {\doibase 10.1103/PhysRevD.108.105023} {\bibfield
  {journal} {\bibinfo  {journal} {Phys. Rev. D}\ }\textbf {\bibinfo {volume}
  {108}},\ \bibinfo {pages} {105023} (\bibinfo {year} {2023})}\BibitemShut
  {NoStop}%
\bibitem [{\citenamefont {Novotn\'y}\ and\ \citenamefont
  {Str\'ansk\'y}(2023)}]{Novotny2023}%
  \BibitemOpen
  \bibfield  {author} {\bibinfo {author} {\bibfnamefont {J.}~\bibnamefont
  {Novotn\'y}}\ and\ \bibinfo {author} {\bibfnamefont {P.}~\bibnamefont
  {Str\'ansk\'y}},\ }\href {\doibase 10.1103/PhysRevE.107.054220} {\bibfield
  {journal} {\bibinfo  {journal} {Phys. Rev. E}\ }\textbf {\bibinfo {volume}
  {107}},\ \bibinfo {pages} {054220} (\bibinfo {year} {2023})}\BibitemShut
  {NoStop}%
\bibitem [{\citenamefont {García-Mata}\ \emph {et~al.}(2023)\citenamefont
  {García-Mata}, \citenamefont {Jalabert},\ and\ \citenamefont
  {Wisniacki}}]{Garcia2023}%
  \BibitemOpen
  \bibfield  {author} {\bibinfo {author} {\bibfnamefont {I.}~\bibnamefont
  {García-Mata}}, \bibinfo {author} {\bibfnamefont {R.~A.}\ \bibnamefont
  {Jalabert}}, \ and\ \bibinfo {author} {\bibfnamefont {D.~A.}\ \bibnamefont
  {Wisniacki}},\ }\href {\doibase 10.4249/scholarpedia.55237} {\bibfield
  {journal} {\bibinfo  {journal} {Scholarpedia}\ }\textbf {\bibinfo {volume}
  {18}},\ \bibinfo {pages} {55237} (\bibinfo {year} {2023})},\ \bibinfo {note}
  {revision \#199677}\BibitemShut {NoStop}%
\bibitem [{\citenamefont {Shukla}\ \emph {et~al.}(2024)\citenamefont {Shukla},
  \citenamefont {Malik}, \citenamefont {Aravinda},\ and\ \citenamefont
  {Mishra}}]{Shukla2024}%
  \BibitemOpen
  \bibfield  {author} {\bibinfo {author} {\bibfnamefont {R.~K.}\ \bibnamefont
  {Shukla}}, \bibinfo {author} {\bibfnamefont {G.~R.}\ \bibnamefont {Malik}},
  \bibinfo {author} {\bibfnamefont {S.}~\bibnamefont {Aravinda}}, \ and\
  \bibinfo {author} {\bibfnamefont {S.~K.}\ \bibnamefont {Mishra}},\ }\href
  {https://arxiv.org/abs/2404.04177} {\enquote {\bibinfo {title}
  {Discriminating chaotic and integrable regimes in quenched field floquet
  system using saturation of out-of-time-order correlation},}\ } (\bibinfo
  {year} {2024}),\ \Eprint {http://arxiv.org/abs/2404.04177} {arXiv:2404.04177
  [quant-ph]} \BibitemShut {NoStop}%
\bibitem [{\citenamefont {Li}\ \emph {et~al.}(2017)\citenamefont {Li},
  \citenamefont {Fan}, \citenamefont {Wang}, \citenamefont {Ye}, \citenamefont
  {Zeng}, \citenamefont {Zhai}, \citenamefont {Peng},\ and\ \citenamefont
  {Du}}]{LiJun2017}%
  \BibitemOpen
  \bibfield  {author} {\bibinfo {author} {\bibfnamefont {J.}~\bibnamefont
  {Li}}, \bibinfo {author} {\bibfnamefont {R.}~\bibnamefont {Fan}}, \bibinfo
  {author} {\bibfnamefont {H.}~\bibnamefont {Wang}}, \bibinfo {author}
  {\bibfnamefont {B.}~\bibnamefont {Ye}}, \bibinfo {author} {\bibfnamefont
  {B.}~\bibnamefont {Zeng}}, \bibinfo {author} {\bibfnamefont {H.}~\bibnamefont
  {Zhai}}, \bibinfo {author} {\bibfnamefont {X.}~\bibnamefont {Peng}}, \ and\
  \bibinfo {author} {\bibfnamefont {J.}~\bibnamefont {Du}},\ }\href {\doibase
  10.1103/PhysRevX.7.031011} {\bibfield  {journal} {\bibinfo  {journal} {Phys.
  Rev. X}\ }\textbf {\bibinfo {volume} {7}},\ \bibinfo {pages} {031011}
  (\bibinfo {year} {2017})}\BibitemShut {NoStop}%
\bibitem [{\citenamefont {G{\"a}rttner}\ \emph {et~al.}(2017)\citenamefont
  {G{\"a}rttner}, \citenamefont {Bohnet}, \citenamefont {Safavi-Naini},
  \citenamefont {Wall}, \citenamefont {Bollinger},\ and\ \citenamefont
  {Rey}}]{Garttner2017}%
  \BibitemOpen
  \bibfield  {author} {\bibinfo {author} {\bibfnamefont {M.}~\bibnamefont
  {G{\"a}rttner}}, \bibinfo {author} {\bibfnamefont {J.~G.}\ \bibnamefont
  {Bohnet}}, \bibinfo {author} {\bibfnamefont {A.}~\bibnamefont
  {Safavi-Naini}}, \bibinfo {author} {\bibfnamefont {M.~L.}\ \bibnamefont
  {Wall}}, \bibinfo {author} {\bibfnamefont {J.~J.}\ \bibnamefont {Bollinger}},
  \ and\ \bibinfo {author} {\bibfnamefont {A.~M.}\ \bibnamefont {Rey}},\ }\href
  {\doibase 10.1038/nphys4119} {\bibfield  {journal} {\bibinfo  {journal}
  {Nature Physics}\ }\textbf {\bibinfo {volume} {13}},\ \bibinfo {pages} {781}
  (\bibinfo {year} {2017})}\BibitemShut {NoStop}%
\bibitem [{\citenamefont {Braum{\"u}ller}\ \emph {et~al.}(2022)\citenamefont
  {Braum{\"u}ller}, \citenamefont {Karamlou}, \citenamefont {Yanay},
  \citenamefont {Kannan}, \citenamefont {Kim}, \citenamefont {Kjaergaard},
  \citenamefont {Melville}, \citenamefont {Niedzielski}, \citenamefont {Sung},
  \citenamefont {Veps{\"a}l{\"a}inen}, \citenamefont {Winik}, \citenamefont
  {Yoder}, \citenamefont {Orlando}, \citenamefont {Gustavsson}, \citenamefont
  {Tahan},\ and\ \citenamefont {Oliver}}]{Braumuller2022}%
  \BibitemOpen
  \bibfield  {author} {\bibinfo {author} {\bibfnamefont {J.}~\bibnamefont
  {Braum{\"u}ller}}, \bibinfo {author} {\bibfnamefont {A.~H.}\ \bibnamefont
  {Karamlou}}, \bibinfo {author} {\bibfnamefont {Y.}~\bibnamefont {Yanay}},
  \bibinfo {author} {\bibfnamefont {B.}~\bibnamefont {Kannan}}, \bibinfo
  {author} {\bibfnamefont {D.}~\bibnamefont {Kim}}, \bibinfo {author}
  {\bibfnamefont {M.}~\bibnamefont {Kjaergaard}}, \bibinfo {author}
  {\bibfnamefont {A.}~\bibnamefont {Melville}}, \bibinfo {author}
  {\bibfnamefont {B.~M.}\ \bibnamefont {Niedzielski}}, \bibinfo {author}
  {\bibfnamefont {Y.}~\bibnamefont {Sung}}, \bibinfo {author} {\bibfnamefont
  {A.}~\bibnamefont {Veps{\"a}l{\"a}inen}}, \bibinfo {author} {\bibfnamefont
  {R.}~\bibnamefont {Winik}}, \bibinfo {author} {\bibfnamefont {J.~L.}\
  \bibnamefont {Yoder}}, \bibinfo {author} {\bibfnamefont {T.~P.}\ \bibnamefont
  {Orlando}}, \bibinfo {author} {\bibfnamefont {S.}~\bibnamefont {Gustavsson}},
  \bibinfo {author} {\bibfnamefont {C.}~\bibnamefont {Tahan}}, \ and\ \bibinfo
  {author} {\bibfnamefont {W.~D.}\ \bibnamefont {Oliver}},\ }\href {\doibase
  10.1038/s41567-021-01430-w} {\bibfield  {journal} {\bibinfo  {journal}
  {Nature Physics}\ }\textbf {\bibinfo {volume} {18}},\ \bibinfo {pages} {172}
  (\bibinfo {year} {2022})}\BibitemShut {NoStop}%
\bibitem [{\citenamefont {Green}\ \emph {et~al.}(2022)\citenamefont {Green},
  \citenamefont {Elben}, \citenamefont {Alderete}, \citenamefont {Joshi},
  \citenamefont {Nguyen}, \citenamefont {Zache}, \citenamefont {Zhu},
  \citenamefont {Sundar},\ and\ \citenamefont {Linke}}]{Green2022}%
  \BibitemOpen
  \bibfield  {author} {\bibinfo {author} {\bibfnamefont {A.~M.}\ \bibnamefont
  {Green}}, \bibinfo {author} {\bibfnamefont {A.}~\bibnamefont {Elben}},
  \bibinfo {author} {\bibfnamefont {C.~H.}\ \bibnamefont {Alderete}}, \bibinfo
  {author} {\bibfnamefont {L.~K.}\ \bibnamefont {Joshi}}, \bibinfo {author}
  {\bibfnamefont {N.~H.}\ \bibnamefont {Nguyen}}, \bibinfo {author}
  {\bibfnamefont {T.~V.}\ \bibnamefont {Zache}}, \bibinfo {author}
  {\bibfnamefont {Y.}~\bibnamefont {Zhu}}, \bibinfo {author} {\bibfnamefont
  {B.}~\bibnamefont {Sundar}}, \ and\ \bibinfo {author} {\bibfnamefont {N.~M.}\
  \bibnamefont {Linke}},\ }\href {\doibase 10.1103/PhysRevLett.128.140601}
  {\bibfield  {journal} {\bibinfo  {journal} {Phys. Rev. Lett.}\ }\textbf
  {\bibinfo {volume} {128}},\ \bibinfo {pages} {140601} (\bibinfo {year}
  {2022})}\BibitemShut {NoStop}%
\bibitem [{\citenamefont {Blocher}\ \emph {et~al.}(2022)\citenamefont
  {Blocher}, \citenamefont {Asaad}, \citenamefont {Mourik}, \citenamefont
  {Johnson}, \citenamefont {Morello},\ and\ \citenamefont
  {M\o{}lmer}}]{Blocher2022}%
  \BibitemOpen
  \bibfield  {author} {\bibinfo {author} {\bibfnamefont {P.~D.}\ \bibnamefont
  {Blocher}}, \bibinfo {author} {\bibfnamefont {S.}~\bibnamefont {Asaad}},
  \bibinfo {author} {\bibfnamefont {V.}~\bibnamefont {Mourik}}, \bibinfo
  {author} {\bibfnamefont {M.~A.~I.}\ \bibnamefont {Johnson}}, \bibinfo
  {author} {\bibfnamefont {A.}~\bibnamefont {Morello}}, \ and\ \bibinfo
  {author} {\bibfnamefont {K.}~\bibnamefont {M\o{}lmer}},\ }\href {\doibase
  10.1103/PhysRevA.106.042429} {\bibfield  {journal} {\bibinfo  {journal}
  {Phys. Rev. A}\ }\textbf {\bibinfo {volume} {106}},\ \bibinfo {pages}
  {042429} (\bibinfo {year} {2022})}\BibitemShut {NoStop}%
\bibitem [{\citenamefont {Lashkari}\ \emph {et~al.}(2013)\citenamefont
  {Lashkari}, \citenamefont {Stanford}, \citenamefont {Hastings}, \citenamefont
  {Osborne},\ and\ \citenamefont {Hayden}}]{Lashkari2013}%
  \BibitemOpen
  \bibfield  {author} {\bibinfo {author} {\bibfnamefont {N.}~\bibnamefont
  {Lashkari}}, \bibinfo {author} {\bibfnamefont {D.}~\bibnamefont {Stanford}},
  \bibinfo {author} {\bibfnamefont {M.}~\bibnamefont {Hastings}}, \bibinfo
  {author} {\bibfnamefont {T.}~\bibnamefont {Osborne}}, \ and\ \bibinfo
  {author} {\bibfnamefont {P.}~\bibnamefont {Hayden}},\ }\href {\doibase
  10.1007/JHEP04(2013)022} {\bibfield  {journal} {\bibinfo  {journal} {Journal
  of High Energy Physics}\ }\textbf {\bibinfo {volume} {2013}},\ \bibinfo
  {pages} {22} (\bibinfo {year} {2013})}\BibitemShut {NoStop}%
\bibitem [{\citenamefont {Kukuljan}\ \emph {et~al.}(2017)\citenamefont
  {Kukuljan}, \citenamefont {Grozdanov},\ and\ \citenamefont
  {Prosen}}]{Kukuljan2017}%
  \BibitemOpen
  \bibfield  {author} {\bibinfo {author} {\bibfnamefont {I.}~\bibnamefont
  {Kukuljan}}, \bibinfo {author} {\bibfnamefont {S.}~\bibnamefont {Grozdanov}},
  \ and\ \bibinfo {author} {\bibfnamefont {T.}~\bibnamefont {Prosen}},\ }\href
  {\doibase 10.1103/PhysRevB.96.060301} {\bibfield  {journal} {\bibinfo
  {journal} {Phys. Rev. B}\ }\textbf {\bibinfo {volume} {96}},\ \bibinfo
  {pages} {060301} (\bibinfo {year} {2017})}\BibitemShut {NoStop}%
\bibitem [{\citenamefont {Schuster}\ \emph {et~al.}(2023)\citenamefont
  {Schuster}, \citenamefont {Niu}, \citenamefont {Cotler}, \citenamefont
  {O'Brien}, \citenamefont {McClean},\ and\ \citenamefont
  {Mohseni}}]{Schuster2023}%
  \BibitemOpen
  \bibfield  {author} {\bibinfo {author} {\bibfnamefont {T.}~\bibnamefont
  {Schuster}}, \bibinfo {author} {\bibfnamefont {M.}~\bibnamefont {Niu}},
  \bibinfo {author} {\bibfnamefont {J.}~\bibnamefont {Cotler}}, \bibinfo
  {author} {\bibfnamefont {T.}~\bibnamefont {O'Brien}}, \bibinfo {author}
  {\bibfnamefont {J.~R.}\ \bibnamefont {McClean}}, \ and\ \bibinfo {author}
  {\bibfnamefont {M.}~\bibnamefont {Mohseni}},\ }\href {\doibase
  10.1103/PhysRevResearch.5.043284} {\bibfield  {journal} {\bibinfo  {journal}
  {Phys. Rev. Res.}\ }\textbf {\bibinfo {volume} {5}},\ \bibinfo {pages}
  {043284} (\bibinfo {year} {2023})}\BibitemShut {NoStop}%
\bibitem [{\citenamefont {Hashimoto}\ \emph {et~al.}(2017)\citenamefont
  {Hashimoto}, \citenamefont {Murata},\ and\ \citenamefont
  {Yoshii}}]{Hashimoto2017}%
  \BibitemOpen
  \bibfield  {author} {\bibinfo {author} {\bibfnamefont {K.}~\bibnamefont
  {Hashimoto}}, \bibinfo {author} {\bibfnamefont {K.}~\bibnamefont {Murata}}, \
  and\ \bibinfo {author} {\bibfnamefont {R.}~\bibnamefont {Yoshii}},\ }\href
  {\doibase 10.1007/JHEP10(2017)138} {\bibfield  {journal} {\bibinfo  {journal}
  {Journal of High Energy Physics}\ }\textbf {\bibinfo {volume} {2017}},\
  \bibinfo {pages} {138} (\bibinfo {year} {2017})}\BibitemShut {NoStop}%
\bibitem [{\citenamefont {Yan}\ \emph {et~al.}(2019)\citenamefont {Yan},
  \citenamefont {Wang},\ and\ \citenamefont {Wang}}]{Yan2019}%
  \BibitemOpen
  \bibfield  {author} {\bibinfo {author} {\bibfnamefont {H.}~\bibnamefont
  {Yan}}, \bibinfo {author} {\bibfnamefont {J.-Z.}\ \bibnamefont {Wang}}, \
  and\ \bibinfo {author} {\bibfnamefont {W.-G.}\ \bibnamefont {Wang}},\ }\href
  {\doibase 10.1088/0253-6102/71/11/1359} {\bibfield  {journal} {\bibinfo
  {journal} {Communications in Theoretical Physics}\ }\textbf {\bibinfo
  {volume} {71}},\ \bibinfo {pages} {1359} (\bibinfo {year}
  {2019})}\BibitemShut {NoStop}%
\bibitem [{\citenamefont {Pilatowsky-Cameo}\ \emph {et~al.}(2020)\citenamefont
  {Pilatowsky-Cameo}, \citenamefont {Ch\'avez-Carlos}, \citenamefont
  {Bastarrachea-Magnani}, \citenamefont {Str\'ansk\'y}, \citenamefont
  {Lerma-Hern\'andez}, \citenamefont {Santos},\ and\ \citenamefont
  {Hirsch}}]{Cameo2020}%
  \BibitemOpen
  \bibfield  {author} {\bibinfo {author} {\bibfnamefont {S.}~\bibnamefont
  {Pilatowsky-Cameo}}, \bibinfo {author} {\bibfnamefont {J.}~\bibnamefont
  {Ch\'avez-Carlos}}, \bibinfo {author} {\bibfnamefont {M.~A.}\ \bibnamefont
  {Bastarrachea-Magnani}}, \bibinfo {author} {\bibfnamefont {P.}~\bibnamefont
  {Str\'ansk\'y}}, \bibinfo {author} {\bibfnamefont {S.}~\bibnamefont
  {Lerma-Hern\'andez}}, \bibinfo {author} {\bibfnamefont {L.~F.}\ \bibnamefont
  {Santos}}, \ and\ \bibinfo {author} {\bibfnamefont {J.~G.}\ \bibnamefont
  {Hirsch}},\ }\href {\doibase 10.1103/PhysRevE.101.010202} {\bibfield
  {journal} {\bibinfo  {journal} {Phys. Rev. E}\ }\textbf {\bibinfo {volume}
  {101}},\ \bibinfo {pages} {010202} (\bibinfo {year} {2020})}\BibitemShut
  {NoStop}%
\bibitem [{\citenamefont {Rozenbaum}\ \emph {et~al.}(2020)\citenamefont
  {Rozenbaum}, \citenamefont {Bunimovich},\ and\ \citenamefont
  {Galitski}}]{Rozenbaum2020}%
  \BibitemOpen
  \bibfield  {author} {\bibinfo {author} {\bibfnamefont {E.~B.}\ \bibnamefont
  {Rozenbaum}}, \bibinfo {author} {\bibfnamefont {L.~A.}\ \bibnamefont
  {Bunimovich}}, \ and\ \bibinfo {author} {\bibfnamefont {V.}~\bibnamefont
  {Galitski}},\ }\href {\doibase 10.1103/PhysRevLett.125.014101} {\bibfield
  {journal} {\bibinfo  {journal} {Phys. Rev. Lett.}\ }\textbf {\bibinfo
  {volume} {125}},\ \bibinfo {pages} {014101} (\bibinfo {year}
  {2020})}\BibitemShut {NoStop}%
\bibitem [{\citenamefont {Hashimoto}\ \emph {et~al.}(2020)\citenamefont
  {Hashimoto}, \citenamefont {Huh}, \citenamefont {Kim},\ and\ \citenamefont
  {Watanabe}}]{Hashimoto2020}%
  \BibitemOpen
  \bibfield  {author} {\bibinfo {author} {\bibfnamefont {K.}~\bibnamefont
  {Hashimoto}}, \bibinfo {author} {\bibfnamefont {K.-B.}\ \bibnamefont {Huh}},
  \bibinfo {author} {\bibfnamefont {K.-Y.}\ \bibnamefont {Kim}}, \ and\
  \bibinfo {author} {\bibfnamefont {R.}~\bibnamefont {Watanabe}},\ }\href
  {\doibase 10.1007/JHEP11(2020)068} {\bibfield  {journal} {\bibinfo  {journal}
  {Journal of High Energy Physics}\ }\textbf {\bibinfo {volume} {2020}},\
  \bibinfo {pages} {68} (\bibinfo {year} {2020})}\BibitemShut {NoStop}%
\bibitem [{\citenamefont {Wang}\ \emph {et~al.}(2021)\citenamefont {Wang},
  \citenamefont {Benenti}, \citenamefont {Casati},\ and\ \citenamefont
  {Wang}}]{WangJ2021}%
  \BibitemOpen
  \bibfield  {author} {\bibinfo {author} {\bibfnamefont {J.}~\bibnamefont
  {Wang}}, \bibinfo {author} {\bibfnamefont {G.}~\bibnamefont {Benenti}},
  \bibinfo {author} {\bibfnamefont {G.}~\bibnamefont {Casati}}, \ and\ \bibinfo
  {author} {\bibfnamefont {W.-g.}\ \bibnamefont {Wang}},\ }\href {\doibase
  10.1103/PhysRevE.103.L030201} {\bibfield  {journal} {\bibinfo  {journal}
  {Phys. Rev. E}\ }\textbf {\bibinfo {volume} {103}},\ \bibinfo {pages}
  {L030201} (\bibinfo {year} {2021})}\BibitemShut {NoStop}%
\bibitem [{\citenamefont {Lakshminarayan}(2019)}]{Lakshmin2019}%
  \BibitemOpen
  \bibfield  {author} {\bibinfo {author} {\bibfnamefont {A.}~\bibnamefont
  {Lakshminarayan}},\ }\href {\doibase 10.1103/PhysRevE.99.012201} {\bibfield
  {journal} {\bibinfo  {journal} {Phys. Rev. E}\ }\textbf {\bibinfo {volume}
  {99}},\ \bibinfo {pages} {012201} (\bibinfo {year} {2019})}\BibitemShut
  {NoStop}%
\bibitem [{\citenamefont {Torres-Herrera}\ \emph {et~al.}(2018)\citenamefont
  {Torres-Herrera}, \citenamefont {Garc\'{\i}a-Garc\'{\i}a},\ and\
  \citenamefont {Santos}}]{Herrera2018}%
  \BibitemOpen
  \bibfield  {author} {\bibinfo {author} {\bibfnamefont {E.~J.}\ \bibnamefont
  {Torres-Herrera}}, \bibinfo {author} {\bibfnamefont {A.~M.}\ \bibnamefont
  {Garc\'{\i}a-Garc\'{\i}a}}, \ and\ \bibinfo {author} {\bibfnamefont {L.~F.}\
  \bibnamefont {Santos}},\ }\href {\doibase 10.1103/PhysRevB.97.060303}
  {\bibfield  {journal} {\bibinfo  {journal} {Phys. Rev. B}\ }\textbf {\bibinfo
  {volume} {97}},\ \bibinfo {pages} {060303} (\bibinfo {year}
  {2018})}\BibitemShut {NoStop}%
\bibitem [{\citenamefont {Rammensee}\ \emph {et~al.}(2018)\citenamefont
  {Rammensee}, \citenamefont {Urbina},\ and\ \citenamefont
  {Richter}}]{Rammensee2018}%
  \BibitemOpen
  \bibfield  {author} {\bibinfo {author} {\bibfnamefont {J.}~\bibnamefont
  {Rammensee}}, \bibinfo {author} {\bibfnamefont {J.~D.}\ \bibnamefont
  {Urbina}}, \ and\ \bibinfo {author} {\bibfnamefont {K.}~\bibnamefont
  {Richter}},\ }\href {\doibase 10.1103/PhysRevLett.121.124101} {\bibfield
  {journal} {\bibinfo  {journal} {Phys. Rev. Lett.}\ }\textbf {\bibinfo
  {volume} {121}},\ \bibinfo {pages} {124101} (\bibinfo {year}
  {2018})}\BibitemShut {NoStop}%
\bibitem [{\citenamefont {Bergamasco}\ \emph {et~al.}(2019)\citenamefont
  {Bergamasco}, \citenamefont {Carlo},\ and\ \citenamefont
  {Rivas}}]{Bergamasco2019}%
  \BibitemOpen
  \bibfield  {author} {\bibinfo {author} {\bibfnamefont {P.~D.}\ \bibnamefont
  {Bergamasco}}, \bibinfo {author} {\bibfnamefont {G.~G.}\ \bibnamefont
  {Carlo}}, \ and\ \bibinfo {author} {\bibfnamefont {A.~M.~F.}\ \bibnamefont
  {Rivas}},\ }\href {\doibase 10.1103/PhysRevResearch.1.033044} {\bibfield
  {journal} {\bibinfo  {journal} {Phys. Rev. Res.}\ }\textbf {\bibinfo {volume}
  {1}},\ \bibinfo {pages} {033044} (\bibinfo {year} {2019})}\BibitemShut
  {NoStop}%
\bibitem [{\citenamefont {Borgonovi}\ \emph {et~al.}(2019)\citenamefont
  {Borgonovi}, \citenamefont {Izrailev},\ and\ \citenamefont
  {Santos}}]{Borgonovi2019}%
  \BibitemOpen
  \bibfield  {author} {\bibinfo {author} {\bibfnamefont {F.}~\bibnamefont
  {Borgonovi}}, \bibinfo {author} {\bibfnamefont {F.~M.}\ \bibnamefont
  {Izrailev}}, \ and\ \bibinfo {author} {\bibfnamefont {L.~F.}\ \bibnamefont
  {Santos}},\ }\href {\doibase 10.1103/PhysRevE.99.052143} {\bibfield
  {journal} {\bibinfo  {journal} {Phys. Rev. E}\ }\textbf {\bibinfo {volume}
  {99}},\ \bibinfo {pages} {052143} (\bibinfo {year} {2019})}\BibitemShut
  {NoStop}%
\bibitem [{\citenamefont {Riddell}\ \emph {et~al.}(2023)\citenamefont
  {Riddell}, \citenamefont {Kirkby}, \citenamefont {O'Dell},\ and\
  \citenamefont {S\o{}rensen}}]{Riddell2023}%
  \BibitemOpen
  \bibfield  {author} {\bibinfo {author} {\bibfnamefont {J.}~\bibnamefont
  {Riddell}}, \bibinfo {author} {\bibfnamefont {W.}~\bibnamefont {Kirkby}},
  \bibinfo {author} {\bibfnamefont {D.~H.~J.}\ \bibnamefont {O'Dell}}, \ and\
  \bibinfo {author} {\bibfnamefont {E.~S.}\ \bibnamefont {S\o{}rensen}},\
  }\href {\doibase 10.1103/PhysRevB.108.L121108} {\bibfield  {journal}
  {\bibinfo  {journal} {Phys. Rev. B}\ }\textbf {\bibinfo {volume} {108}},\
  \bibinfo {pages} {L121108} (\bibinfo {year} {2023})}\BibitemShut {NoStop}%
\bibitem [{\citenamefont {Balachandran}\ \emph {et~al.}(2023)\citenamefont
  {Balachandran}, \citenamefont {Santos}, \citenamefont {Rigol},\ and\
  \citenamefont {Poletti}}]{Balachandran2023}%
  \BibitemOpen
  \bibfield  {author} {\bibinfo {author} {\bibfnamefont {V.}~\bibnamefont
  {Balachandran}}, \bibinfo {author} {\bibfnamefont {L.~F.}\ \bibnamefont
  {Santos}}, \bibinfo {author} {\bibfnamefont {M.}~\bibnamefont {Rigol}}, \
  and\ \bibinfo {author} {\bibfnamefont {D.}~\bibnamefont {Poletti}},\ }\href
  {\doibase 10.1103/PhysRevB.107.235421} {\bibfield  {journal} {\bibinfo
  {journal} {Phys. Rev. B}\ }\textbf {\bibinfo {volume} {107}},\ \bibinfo
  {pages} {235421} (\bibinfo {year} {2023})}\BibitemShut {NoStop}%
\bibitem [{\citenamefont {Varikuti}\ and\ \citenamefont
  {Madhok}(2024)}]{Varikuti2024}%
  \BibitemOpen
  \bibfield  {author} {\bibinfo {author} {\bibfnamefont {N.~D.}\ \bibnamefont
  {Varikuti}}\ and\ \bibinfo {author} {\bibfnamefont {V.}~\bibnamefont
  {Madhok}},\ }\href {\doibase 10.1063/5.0191140} {\bibfield  {journal}
  {\bibinfo  {journal} {Chaos: An Interdisciplinary Journal of Nonlinear
  Science}\ }\textbf {\bibinfo {volume} {34}},\ \bibinfo {pages} {063124}
  (\bibinfo {year} {2024})}\BibitemShut {NoStop}%
\bibitem [{\citenamefont {Syzranov}\ \emph {et~al.}(2018)\citenamefont
  {Syzranov}, \citenamefont {Gorshkov},\ and\ \citenamefont
  {Galitski}}]{Syzranov2018}%
  \BibitemOpen
  \bibfield  {author} {\bibinfo {author} {\bibfnamefont {S.~V.}\ \bibnamefont
  {Syzranov}}, \bibinfo {author} {\bibfnamefont {A.~V.}\ \bibnamefont
  {Gorshkov}}, \ and\ \bibinfo {author} {\bibfnamefont {V.}~\bibnamefont
  {Galitski}},\ }\href {\doibase 10.1103/PhysRevB.97.161114} {\bibfield
  {journal} {\bibinfo  {journal} {Phys. Rev. B}\ }\textbf {\bibinfo {volume}
  {97}},\ \bibinfo {pages} {161114} (\bibinfo {year} {2018})}\BibitemShut
  {NoStop}%
\bibitem [{\citenamefont {Chatterjee}\ \emph {et~al.}(2020)\citenamefont
  {Chatterjee}, \citenamefont {Kundu},\ and\ \citenamefont
  {Kulkarni}}]{Chatterjee2020}%
  \BibitemOpen
  \bibfield  {author} {\bibinfo {author} {\bibfnamefont {A.~K.}\ \bibnamefont
  {Chatterjee}}, \bibinfo {author} {\bibfnamefont {A.}~\bibnamefont {Kundu}}, \
  and\ \bibinfo {author} {\bibfnamefont {M.}~\bibnamefont {Kulkarni}},\ }\href
  {\doibase 10.1103/PhysRevE.102.052103} {\bibfield  {journal} {\bibinfo
  {journal} {Phys. Rev. E}\ }\textbf {\bibinfo {volume} {102}},\ \bibinfo
  {pages} {052103} (\bibinfo {year} {2020})}\BibitemShut {NoStop}%
\bibitem [{\citenamefont {Zhai}\ and\ \citenamefont {Yin}(2020)}]{ZhaiL2020}%
  \BibitemOpen
  \bibfield  {author} {\bibinfo {author} {\bibfnamefont {L.-J.}\ \bibnamefont
  {Zhai}}\ and\ \bibinfo {author} {\bibfnamefont {S.}~\bibnamefont {Yin}},\
  }\href {\doibase 10.1103/PhysRevB.102.054303} {\bibfield  {journal} {\bibinfo
   {journal} {Phys. Rev. B}\ }\textbf {\bibinfo {volume} {102}},\ \bibinfo
  {pages} {054303} (\bibinfo {year} {2020})}\BibitemShut {NoStop}%
\bibitem [{\citenamefont {Zhao}(2022)}]{ZhaoW2022}%
  \BibitemOpen
  \bibfield  {author} {\bibinfo {author} {\bibfnamefont {W.-L.}\ \bibnamefont
  {Zhao}},\ }\href {\doibase 10.1103/PhysRevResearch.4.023004} {\bibfield
  {journal} {\bibinfo  {journal} {Phys. Rev. Res.}\ }\textbf {\bibinfo {volume}
  {4}},\ \bibinfo {pages} {023004} (\bibinfo {year} {2022})}\BibitemShut
  {NoStop}%
\bibitem [{\citenamefont {Bergamasco}\ \emph {et~al.}(2023)\citenamefont
  {Bergamasco}, \citenamefont {Carlo},\ and\ \citenamefont
  {Rivas}}]{Bergamasco2023}%
  \BibitemOpen
  \bibfield  {author} {\bibinfo {author} {\bibfnamefont {P.~D.}\ \bibnamefont
  {Bergamasco}}, \bibinfo {author} {\bibfnamefont {G.~G.}\ \bibnamefont
  {Carlo}}, \ and\ \bibinfo {author} {\bibfnamefont {A.~M.~F.}\ \bibnamefont
  {Rivas}},\ }\href {\doibase 10.1103/PhysRevE.108.024208} {\bibfield
  {journal} {\bibinfo  {journal} {Phys. Rev. E}\ }\textbf {\bibinfo {volume}
  {108}},\ \bibinfo {pages} {024208} (\bibinfo {year} {2023})}\BibitemShut
  {NoStop}%
\bibitem [{\citenamefont {Kidd}\ \emph {et~al.}(2021)\citenamefont {Kidd},
  \citenamefont {Safavi-Naini},\ and\ \citenamefont {Corney}}]{Kidd2021}%
  \BibitemOpen
  \bibfield  {author} {\bibinfo {author} {\bibfnamefont {R.~A.}\ \bibnamefont
  {Kidd}}, \bibinfo {author} {\bibfnamefont {A.}~\bibnamefont {Safavi-Naini}},
  \ and\ \bibinfo {author} {\bibfnamefont {J.~F.}\ \bibnamefont {Corney}},\
  }\href {\doibase 10.1103/PhysRevA.103.033304} {\bibfield  {journal} {\bibinfo
   {journal} {Phys. Rev. A}\ }\textbf {\bibinfo {volume} {103}},\ \bibinfo
  {pages} {033304} (\bibinfo {year} {2021})}\BibitemShut {NoStop}%
\bibitem [{\citenamefont {Brenes}\ \emph {et~al.}(2021)\citenamefont {Brenes},
  \citenamefont {Pappalardi}, \citenamefont {Mitchison}, \citenamefont
  {Goold},\ and\ \citenamefont {Silva}}]{Brenes2021}%
  \BibitemOpen
  \bibfield  {author} {\bibinfo {author} {\bibfnamefont {M.}~\bibnamefont
  {Brenes}}, \bibinfo {author} {\bibfnamefont {S.}~\bibnamefont {Pappalardi}},
  \bibinfo {author} {\bibfnamefont {M.~T.}\ \bibnamefont {Mitchison}}, \bibinfo
  {author} {\bibfnamefont {J.}~\bibnamefont {Goold}}, \ and\ \bibinfo {author}
  {\bibfnamefont {A.}~\bibnamefont {Silva}},\ }\href {\doibase
  10.1103/PhysRevE.104.034120} {\bibfield  {journal} {\bibinfo  {journal}
  {Phys. Rev. E}\ }\textbf {\bibinfo {volume} {104}},\ \bibinfo {pages}
  {034120} (\bibinfo {year} {2021})}\BibitemShut {NoStop}%
\bibitem [{\citenamefont {Heyl}\ \emph {et~al.}(2018)\citenamefont {Heyl},
  \citenamefont {Pollmann},\ and\ \citenamefont {D\'ora}}]{Heyl2018}%
  \BibitemOpen
  \bibfield  {author} {\bibinfo {author} {\bibfnamefont {M.}~\bibnamefont
  {Heyl}}, \bibinfo {author} {\bibfnamefont {F.}~\bibnamefont {Pollmann}}, \
  and\ \bibinfo {author} {\bibfnamefont {B.}~\bibnamefont {D\'ora}},\ }\href
  {\doibase 10.1103/PhysRevLett.121.016801} {\bibfield  {journal} {\bibinfo
  {journal} {Phys. Rev. Lett.}\ }\textbf {\bibinfo {volume} {121}},\ \bibinfo
  {pages} {016801} (\bibinfo {year} {2018})}\BibitemShut {NoStop}%
\bibitem [{\citenamefont {Wang}\ and\ \citenamefont
  {P\'erez-Bernal}(2019)}]{WangQ2019}%
  \BibitemOpen
  \bibfield  {author} {\bibinfo {author} {\bibfnamefont {Q.}~\bibnamefont
  {Wang}}\ and\ \bibinfo {author} {\bibfnamefont {F.}~\bibnamefont
  {P\'erez-Bernal}},\ }\href {\doibase 10.1103/PhysRevA.100.062113} {\bibfield
  {journal} {\bibinfo  {journal} {Phys. Rev. A}\ }\textbf {\bibinfo {volume}
  {100}},\ \bibinfo {pages} {062113} (\bibinfo {year} {2019})}\BibitemShut
  {NoStop}%
\bibitem [{\citenamefont {Nie}\ \emph {et~al.}(2020)\citenamefont {Nie},
  \citenamefont {Wei}, \citenamefont {Chen}, \citenamefont {Zhang},
  \citenamefont {Zhao}, \citenamefont {Qiu}, \citenamefont {Tian},
  \citenamefont {Ji}, \citenamefont {Xin}, \citenamefont {Lu},\ and\
  \citenamefont {Li}}]{Nie2020}%
  \BibitemOpen
  \bibfield  {author} {\bibinfo {author} {\bibfnamefont {X.}~\bibnamefont
  {Nie}}, \bibinfo {author} {\bibfnamefont {B.-B.}\ \bibnamefont {Wei}},
  \bibinfo {author} {\bibfnamefont {X.}~\bibnamefont {Chen}}, \bibinfo {author}
  {\bibfnamefont {Z.}~\bibnamefont {Zhang}}, \bibinfo {author} {\bibfnamefont
  {X.}~\bibnamefont {Zhao}}, \bibinfo {author} {\bibfnamefont {C.}~\bibnamefont
  {Qiu}}, \bibinfo {author} {\bibfnamefont {Y.}~\bibnamefont {Tian}}, \bibinfo
  {author} {\bibfnamefont {Y.}~\bibnamefont {Ji}}, \bibinfo {author}
  {\bibfnamefont {T.}~\bibnamefont {Xin}}, \bibinfo {author} {\bibfnamefont
  {D.}~\bibnamefont {Lu}}, \ and\ \bibinfo {author} {\bibfnamefont
  {J.}~\bibnamefont {Li}},\ }\href {\doibase 10.1103/PhysRevLett.124.250601}
  {\bibfield  {journal} {\bibinfo  {journal} {Phys. Rev. Lett.}\ }\textbf
  {\bibinfo {volume} {124}},\ \bibinfo {pages} {250601} (\bibinfo {year}
  {2020})}\BibitemShut {NoStop}%
\bibitem [{\citenamefont {Chen}\ \emph {et~al.}(2020)\citenamefont {Chen},
  \citenamefont {Hou}, \citenamefont {Zhou}, \citenamefont {Qian},
  \citenamefont {Shen},\ and\ \citenamefont {Xu}}]{ChenB2020}%
  \BibitemOpen
  \bibfield  {author} {\bibinfo {author} {\bibfnamefont {B.}~\bibnamefont
  {Chen}}, \bibinfo {author} {\bibfnamefont {X.}~\bibnamefont {Hou}}, \bibinfo
  {author} {\bibfnamefont {F.}~\bibnamefont {Zhou}}, \bibinfo {author}
  {\bibfnamefont {P.}~\bibnamefont {Qian}}, \bibinfo {author} {\bibfnamefont
  {H.}~\bibnamefont {Shen}}, \ and\ \bibinfo {author} {\bibfnamefont
  {N.}~\bibnamefont {Xu}},\ }\href {\doibase 10.1063/5.0004152} {\bibfield
  {journal} {\bibinfo  {journal} {Applied Physics Letters}\ }\textbf {\bibinfo
  {volume} {116}},\ \bibinfo {pages} {194002} (\bibinfo {year}
  {2020})}\BibitemShut {NoStop}%
\bibitem [{\citenamefont {Huh}\ \emph {et~al.}(2021)\citenamefont {Huh},
  \citenamefont {Ikeda}, \citenamefont {Jahnke},\ and\ \citenamefont
  {Kim}}]{HuhK2021}%
  \BibitemOpen
  \bibfield  {author} {\bibinfo {author} {\bibfnamefont {K.-B.}\ \bibnamefont
  {Huh}}, \bibinfo {author} {\bibfnamefont {K.}~\bibnamefont {Ikeda}}, \bibinfo
  {author} {\bibfnamefont {V.}~\bibnamefont {Jahnke}}, \ and\ \bibinfo {author}
  {\bibfnamefont {K.-Y.}\ \bibnamefont {Kim}},\ }\href {\doibase
  10.1103/PhysRevE.104.024136} {\bibfield  {journal} {\bibinfo  {journal}
  {Phys. Rev. E}\ }\textbf {\bibinfo {volume} {104}},\ \bibinfo {pages}
  {024136} (\bibinfo {year} {2021})}\BibitemShut {NoStop}%
\bibitem [{\citenamefont {Zamani}\ \emph {et~al.}(2022)\citenamefont {Zamani},
  \citenamefont {Jafari},\ and\ \citenamefont {Langari}}]{Zamani2022}%
  \BibitemOpen
  \bibfield  {author} {\bibinfo {author} {\bibfnamefont {S.}~\bibnamefont
  {Zamani}}, \bibinfo {author} {\bibfnamefont {R.}~\bibnamefont {Jafari}}, \
  and\ \bibinfo {author} {\bibfnamefont {A.}~\bibnamefont {Langari}},\ }\href
  {\doibase 10.1103/PhysRevB.105.094304} {\bibfield  {journal} {\bibinfo
  {journal} {Phys. Rev. B}\ }\textbf {\bibinfo {volume} {105}},\ \bibinfo
  {pages} {094304} (\bibinfo {year} {2022})}\BibitemShut {NoStop}%
\bibitem [{\citenamefont {Bin}\ \emph {et~al.}(2023)\citenamefont {Bin},
  \citenamefont {Wan}, \citenamefont {Nori}, \citenamefont {Wu},\ and\
  \citenamefont {L\"u}}]{BinQ2023}%
  \BibitemOpen
  \bibfield  {author} {\bibinfo {author} {\bibfnamefont {Q.}~\bibnamefont
  {Bin}}, \bibinfo {author} {\bibfnamefont {L.-L.}\ \bibnamefont {Wan}},
  \bibinfo {author} {\bibfnamefont {F.}~\bibnamefont {Nori}}, \bibinfo {author}
  {\bibfnamefont {Y.}~\bibnamefont {Wu}}, \ and\ \bibinfo {author}
  {\bibfnamefont {X.-Y.}\ \bibnamefont {L\"u}},\ }\href {\doibase
  10.1103/PhysRevB.107.L020202} {\bibfield  {journal} {\bibinfo  {journal}
  {Phys. Rev. B}\ }\textbf {\bibinfo {volume} {107}},\ \bibinfo {pages}
  {L020202} (\bibinfo {year} {2023})}\BibitemShut {NoStop}%
\bibitem [{\citenamefont {Haake}\ \emph {et~al.}(2019)\citenamefont {Haake},
  \citenamefont {Gnutzmann},\ and\ \citenamefont {Ku{\'s}}}]{Haake2019}%
  \BibitemOpen
  \bibfield  {author} {\bibinfo {author} {\bibfnamefont {F.}~\bibnamefont
  {Haake}}, \bibinfo {author} {\bibfnamefont {S.}~\bibnamefont {Gnutzmann}}, \
  and\ \bibinfo {author} {\bibfnamefont {M.}~\bibnamefont {Ku{\'s}}},\ }\href
  {https://books.google.si/books?id=hHO_uAEACAAJ} {\emph {\bibinfo {title}
  {Quantum Signatures of Chaos}}},\ Springer Series in Synergetics\ (\bibinfo
  {publisher} {Springer International Publishing},\ \bibinfo {year}
  {2019})\BibitemShut {NoStop}%
\bibitem [{\citenamefont {Chaudhury}\ \emph {et~al.}(2009)\citenamefont
  {Chaudhury}, \citenamefont {Smith}, \citenamefont {Anderson}, \citenamefont
  {Ghose},\ and\ \citenamefont {Jessen}}]{Chaudhury2009}%
  \BibitemOpen
  \bibfield  {author} {\bibinfo {author} {\bibfnamefont {S.}~\bibnamefont
  {Chaudhury}}, \bibinfo {author} {\bibfnamefont {A.}~\bibnamefont {Smith}},
  \bibinfo {author} {\bibfnamefont {B.~E.}\ \bibnamefont {Anderson}}, \bibinfo
  {author} {\bibfnamefont {S.}~\bibnamefont {Ghose}}, \ and\ \bibinfo {author}
  {\bibfnamefont {P.~S.}\ \bibnamefont {Jessen}},\ }\href {\doibase
  10.1038/nature08396} {\bibfield  {journal} {\bibinfo  {journal} {Nature}\
  }\textbf {\bibinfo {volume} {461}},\ \bibinfo {pages} {768} (\bibinfo {year}
  {2009})}\BibitemShut {NoStop}%
\bibitem [{\citenamefont {Neill}\ \emph {et~al.}(2016)\citenamefont {Neill},
  \citenamefont {Roushan}, \citenamefont {Fang}, \citenamefont {Chen},
  \citenamefont {Kolodrubetz}, \citenamefont {Chen}, \citenamefont {Megrant},
  \citenamefont {Barends}, \citenamefont {Campbell}, \citenamefont {Chiaro},
  \citenamefont {Dunsworth}, \citenamefont {Jeffrey}, \citenamefont {Kelly},
  \citenamefont {Mutus}, \citenamefont {O'Malley}, \citenamefont {Quintana},
  \citenamefont {Sank}, \citenamefont {Vainsencher}, \citenamefont {Wenner},
  \citenamefont {White}, \citenamefont {Polkovnikov},\ and\ \citenamefont
  {Martinis}}]{Neill2016}%
  \BibitemOpen
  \bibfield  {author} {\bibinfo {author} {\bibfnamefont {C.}~\bibnamefont
  {Neill}}, \bibinfo {author} {\bibfnamefont {P.}~\bibnamefont {Roushan}},
  \bibinfo {author} {\bibfnamefont {M.}~\bibnamefont {Fang}}, \bibinfo {author}
  {\bibfnamefont {Y.}~\bibnamefont {Chen}}, \bibinfo {author} {\bibfnamefont
  {M.}~\bibnamefont {Kolodrubetz}}, \bibinfo {author} {\bibfnamefont
  {Z.}~\bibnamefont {Chen}}, \bibinfo {author} {\bibfnamefont {A.}~\bibnamefont
  {Megrant}}, \bibinfo {author} {\bibfnamefont {R.}~\bibnamefont {Barends}},
  \bibinfo {author} {\bibfnamefont {B.}~\bibnamefont {Campbell}}, \bibinfo
  {author} {\bibfnamefont {B.}~\bibnamefont {Chiaro}}, \bibinfo {author}
  {\bibfnamefont {A.}~\bibnamefont {Dunsworth}}, \bibinfo {author}
  {\bibfnamefont {E.}~\bibnamefont {Jeffrey}}, \bibinfo {author} {\bibfnamefont
  {J.}~\bibnamefont {Kelly}}, \bibinfo {author} {\bibfnamefont
  {J.}~\bibnamefont {Mutus}}, \bibinfo {author} {\bibfnamefont {P.~J.~J.}\
  \bibnamefont {O'Malley}}, \bibinfo {author} {\bibfnamefont {C.}~\bibnamefont
  {Quintana}}, \bibinfo {author} {\bibfnamefont {D.}~\bibnamefont {Sank}},
  \bibinfo {author} {\bibfnamefont {A.}~\bibnamefont {Vainsencher}}, \bibinfo
  {author} {\bibfnamefont {J.}~\bibnamefont {Wenner}}, \bibinfo {author}
  {\bibfnamefont {T.~C.}\ \bibnamefont {White}}, \bibinfo {author}
  {\bibfnamefont {A.}~\bibnamefont {Polkovnikov}}, \ and\ \bibinfo {author}
  {\bibfnamefont {J.~M.}\ \bibnamefont {Martinis}},\ }\href {\doibase
  10.1038/nphys3830} {\bibfield  {journal} {\bibinfo  {journal} {Nature
  Physics}\ }\textbf {\bibinfo {volume} {12}},\ \bibinfo {pages} {1037}
  (\bibinfo {year} {2016})}\BibitemShut {NoStop}%
\bibitem [{\citenamefont {Krithika}\ \emph {et~al.}(2019)\citenamefont
  {Krithika}, \citenamefont {Anjusha}, \citenamefont {Bhosale},\ and\
  \citenamefont {Mahesh}}]{Krithika2019}%
  \BibitemOpen
  \bibfield  {author} {\bibinfo {author} {\bibfnamefont {V.~R.}\ \bibnamefont
  {Krithika}}, \bibinfo {author} {\bibfnamefont {V.~S.}\ \bibnamefont
  {Anjusha}}, \bibinfo {author} {\bibfnamefont {U.~T.}\ \bibnamefont
  {Bhosale}}, \ and\ \bibinfo {author} {\bibfnamefont {T.~S.}\ \bibnamefont
  {Mahesh}},\ }\href {\doibase 10.1103/PhysRevE.99.032219} {\bibfield
  {journal} {\bibinfo  {journal} {Phys. Rev. E}\ }\textbf {\bibinfo {volume}
  {99}},\ \bibinfo {pages} {032219} (\bibinfo {year} {2019})}\BibitemShut
  {NoStop}%
\bibitem [{\citenamefont {Meier}\ \emph {et~al.}(2019)\citenamefont {Meier},
  \citenamefont {Ang'ong'a}, \citenamefont {An},\ and\ \citenamefont
  {Gadway}}]{Meier2019}%
  \BibitemOpen
  \bibfield  {author} {\bibinfo {author} {\bibfnamefont {E.~J.}\ \bibnamefont
  {Meier}}, \bibinfo {author} {\bibfnamefont {J.}~\bibnamefont {Ang'ong'a}},
  \bibinfo {author} {\bibfnamefont {F.~A.}\ \bibnamefont {An}}, \ and\ \bibinfo
  {author} {\bibfnamefont {B.}~\bibnamefont {Gadway}},\ }\href {\doibase
  10.1103/PhysRevA.100.013623} {\bibfield  {journal} {\bibinfo  {journal}
  {Phys. Rev. A}\ }\textbf {\bibinfo {volume} {100}},\ \bibinfo {pages}
  {013623} (\bibinfo {year} {2019})}\BibitemShut {NoStop}%
\bibitem [{\citenamefont {Piga}\ \emph {et~al.}(2019)\citenamefont {Piga},
  \citenamefont {Lewenstein},\ and\ \citenamefont {Quach}}]{Piga2019}%
  \BibitemOpen
  \bibfield  {author} {\bibinfo {author} {\bibfnamefont {A.}~\bibnamefont
  {Piga}}, \bibinfo {author} {\bibfnamefont {M.}~\bibnamefont {Lewenstein}}, \
  and\ \bibinfo {author} {\bibfnamefont {J.~Q.}\ \bibnamefont {Quach}},\ }\href
  {\doibase 10.1103/PhysRevE.99.032213} {\bibfield  {journal} {\bibinfo
  {journal} {Phys. Rev. E}\ }\textbf {\bibinfo {volume} {99}},\ \bibinfo
  {pages} {032213} (\bibinfo {year} {2019})}\BibitemShut {NoStop}%
\bibitem [{\citenamefont {Mu\~noz Arias}\ \emph {et~al.}(2021)\citenamefont
  {Mu\~noz Arias}, \citenamefont {Poggi},\ and\ \citenamefont
  {Deutsch}}]{Munoz2021}%
  \BibitemOpen
  \bibfield  {author} {\bibinfo {author} {\bibfnamefont {M.~H.}\ \bibnamefont
  {Mu\~noz Arias}}, \bibinfo {author} {\bibfnamefont {P.~M.}\ \bibnamefont
  {Poggi}}, \ and\ \bibinfo {author} {\bibfnamefont {I.~H.}\ \bibnamefont
  {Deutsch}},\ }\href {\doibase 10.1103/PhysRevE.103.052212} {\bibfield
  {journal} {\bibinfo  {journal} {Phys. Rev. E}\ }\textbf {\bibinfo {volume}
  {103}},\ \bibinfo {pages} {052212} (\bibinfo {year} {2021})}\BibitemShut
  {NoStop}%
\bibitem [{\citenamefont {Wang}\ and\ \citenamefont
  {Robnik}(2023{\natexlab{b}})}]{QwangR2023}%
  \BibitemOpen
  \bibfield  {author} {\bibinfo {author} {\bibfnamefont {Q.}~\bibnamefont
  {Wang}}\ and\ \bibinfo {author} {\bibfnamefont {M.}~\bibnamefont {Robnik}},\
  }\href {\doibase 10.1103/PhysRevE.107.054213} {\bibfield  {journal} {\bibinfo
   {journal} {Phys. Rev. E}\ }\textbf {\bibinfo {volume} {107}},\ \bibinfo
  {pages} {054213} (\bibinfo {year} {2023}{\natexlab{b}})}\BibitemShut
  {NoStop}%
\bibitem [{\citenamefont {Wang}\ and\ \citenamefont
  {Robnik}(2021)}]{WangR2021}%
  \BibitemOpen
  \bibfield  {author} {\bibinfo {author} {\bibfnamefont {Q.}~\bibnamefont
  {Wang}}\ and\ \bibinfo {author} {\bibfnamefont {M.}~\bibnamefont {Robnik}},\
  }\href {\doibase 10.3390/e23101347} {\bibfield  {journal} {\bibinfo
  {journal} {Entropy}\ }\textbf {\bibinfo {volume} {23}} (\bibinfo {year}
  {2021}),\ 10.3390/e23101347}\BibitemShut {NoStop}%
\bibitem [{\citenamefont {Husimi}(1940)}]{Husimi1940}%
  \BibitemOpen
  \bibfield  {author} {\bibinfo {author} {\bibfnamefont {K.}~\bibnamefont
  {Husimi}},\ }\href {\doibase 10.11429/ppmsj1919.22.4_264} {\bibfield
  {journal} {\bibinfo  {journal} {Nippon Sugaku-Buturigakkwai Kizi Dai 3 Ki}\
  }\textbf {\bibinfo {volume} {22}},\ \bibinfo {pages} {264} (\bibinfo {year}
  {1940})}\BibitemShut {NoStop}%
\bibitem [{\citenamefont {Perelomov}(1977)}]{Perelomov1977}%
  \BibitemOpen
  \bibfield  {author} {\bibinfo {author} {\bibfnamefont {A.~M.}\ \bibnamefont
  {Perelomov}},\ }\href {\doibase 10.1070/PU1977v020n09ABEH005459} {\bibfield
  {journal} {\bibinfo  {journal} {Soviet Physics Uspekhi}\ }\textbf {\bibinfo
  {volume} {20}},\ \bibinfo {pages} {703} (\bibinfo {year} {1977})}\BibitemShut
  {NoStop}%
\bibitem [{\citenamefont {Zhang}\ \emph {et~al.}(1990)\citenamefont {Zhang},
  \citenamefont {Feng},\ and\ \citenamefont {Gilmore}}]{ZhangW1990}%
  \BibitemOpen
  \bibfield  {author} {\bibinfo {author} {\bibfnamefont {W.-M.}\ \bibnamefont
  {Zhang}}, \bibinfo {author} {\bibfnamefont {D.~H.}\ \bibnamefont {Feng}}, \
  and\ \bibinfo {author} {\bibfnamefont {R.}~\bibnamefont {Gilmore}},\ }\href
  {\doibase 10.1103/RevModPhys.62.867} {\bibfield  {journal} {\bibinfo
  {journal} {Rev. Mod. Phys.}\ }\textbf {\bibinfo {volume} {62}},\ \bibinfo
  {pages} {867} (\bibinfo {year} {1990})}\BibitemShut {NoStop}%
\bibitem [{\citenamefont {Sneyd}\ \emph {et~al.}(2022)\citenamefont {Sneyd},
  \citenamefont {Fewster},\ and\ \citenamefont {McGillivray}}]{Sneyd2022}%
  \BibitemOpen
  \bibfield  {author} {\bibinfo {author} {\bibfnamefont {J.}~\bibnamefont
  {Sneyd}}, \bibinfo {author} {\bibfnamefont {R.}~\bibnamefont {Fewster}}, \
  and\ \bibinfo {author} {\bibfnamefont {D.}~\bibnamefont {McGillivray}},\
  }\href {https://books.google.si/books?id=zqd3EAAAQBAJ} {\emph {\bibinfo
  {title} {Mathematics and Statistics for Science}}}\ (\bibinfo  {publisher}
  {Springer International Publishing},\ \bibinfo {year} {2022})\BibitemShut
  {NoStop}%
\end{thebibliography}

%

\end{document}